\documentclass[11pt, a4paper]{article}
\pdfoutput=1
\usepackage[utf8]{inputenc}
\usepackage[T1]{fontenc}
\usepackage{mlmodern}
\usepackage{textcomp}
\usepackage[american]{babel}
\usepackage{microtype}
\usepackage{csquotes}

\usepackage{eurosym}
\usepackage{xspace}
\usepackage[nottoc]{tocbibind}
\usepackage{authblk}
\usepackage{fancyhdr}
\usepackage{totpages}

\usepackage{setspace}
\usepackage{titlesec}
\usepackage{changepage}

\usepackage{graphicx}
\usepackage{subcaption}
\usepackage{float}
\usepackage{makecell}
\usepackage{multirow}
\usepackage{multicol}
\usepackage{enumitem}
\usepackage{marginnote}
\usepackage[multiple]{footmisc}

\usepackage[usenames, dvipsnames, svgnames, table]{xcolor}

\usepackage[
	backend=bibtex8, citestyle=numeric-comp,
	sorting=none, sortcase=false, sortcites=true,
	giveninits=true, maxnames=99
]{biblatex}

\usepackage{amsmath, amsfonts, amssymb}
\usepackage{mathtools}
\usepackage{mathrsfs}
\usepackage{cancel}
\usepackage{braket}
\usepackage{tensor}
\usepackage{slashed}
\usepackage{siunitx}

\usepackage[
	pdftex, breaklinks=true, linktocpage, colorlinks=true,
	urlcolor=RoyalBlue, linkcolor=RoyalBlue, citecolor=BrickRed
]{hyperref}
\usepackage[all]{hypcap}

\usepackage[type={CC}, modifier={by-nc-nd}, version={4.0}]{doclicense}

\newcommand{\email}[1]{\href{mailto:#1}{\nolinkurl{#1}}}
\newcommand{\emailfoot}[1]{\thanks{\email{#1}}}

\newcounter{draftcommentcnt}
\NewDocumentCommand{\draftcomment}{s O{red} m}{%
	\def\margnote{\IfBooleanTF{#1}{\marginnote}{\marginpar}}%
	\stepcounter{draftcommentcnt}%
	\textcolor{#2}{#3}%
	\margnote{\textcolor{#2}{$\Leftarrow$ \arabic{draftcommentcnt}}}%
}

\graphicspath{{./images/}}
\numberwithin{equation}{section}

\usepackage[sort&compress, english]{cleveref}

\usepackage[heightrounded,
	top=3.5cm, bottom=3cm, left=3.5cm, right=3.5cm,
]{geometry}

\newcommand{\e}[0]{\mathrm{e}}
\newcommand{\I}[0]{\mathrm{i}}

\newcommand{\R}[0]{\mathbb{R}}
\newcommand{\C}[0]{\mathbb{C}}
\newcommand{\dd}[0]{\mathrm{d}}

\newcommand{\pd}[0]{\partial}
\newcommand{\mc}[1]{{\mathcal{#1}}}

\renewcommand{\Re}{\operatorname{Re}}
\renewcommand{\Im}{\operatorname{Im}}

\DeclareMathOperator{\Span}{Span}

\newcommand{\abs}[1]{{|#1|}}

\newcommand{\group}[1]{\mathrm{#1}}

\input{arxiv.def}

\hypersetup{
	pdftitle={SL(2, \C) quartic vertex for closed string field theory},
	pdfauthor={Harold Erbin, Suvajit Majumder}
}

\title{$\mathrm{SL}(2, \C)$ quartic vertex for closed string field theory}

\author[1,2,3]{Harold Erbin\emailfoot{erbin@mit.edu}}
\author[4]{Suvajit Majumder\emailfoot{majumder.suvajit95@gmail.com}}

\affil[1]{%
	Université Paris-Saclay, CEA, LIST, F-91120 Palaiseau, France
}

\affil[2]{%
	Center for Theoretical Physics, Massachusetts Institute of Technology
	\protect\\
	Cambridge, MA 02139, USA
}

\affil[3]{%
	NSF AI Institute for Artificial Intelligence and Fundamental Interactions
}

\affil[4]{%
	City, University of London, UK
}

\begin{document}

\maketitle

\begin{abstract}
We construct the $\mathrm{SL}(2, \C)$ quartic vertex with a generic stub parameter for the bosonic closed string field theory by characterizing the vertex region in the moduli space of 4-punctured sphere, and providing the necessary and sufficient constraints for the local coordinate maps.
While $\mathrm{SL}(2, \C)$ vertices are not known to have a nice geometric recursive construction like the minimal area or hyperbolic vertices, they can be studied analytically which makes them more convenient for simple computations.
In particular, we obtain exact formulas for the parametrization and volume of the vertex region as a function of the stub parameter.
The main objective of having an explicit quartic vertex is to later study its decomposition using auxiliary fields.
\end{abstract}

\newpage

\hrule
\pdfbookmark[1]{\contentsname}{toc}
\tableofcontents
\bigskip
\hrule

\section{Introduction}

In string theory, $g$-loop $n$-point off-shell scattering amplitudes are defined by integrating specific correlation functions from a $2d$ CFT over the moduli space $\mc M_{g,n}$ of genus-$g$ Riemann surfaces with $n$ punctures, together with the data of $n$ local coordinates (holomorphic functions).
In order to interpret the scattering amplitude as a sum of Feynman graphs, one splits the moduli space into two disconnected regions, $\mc F_{g,n}$ and $\mc V_{g,n}$.
The region $\mc F_{g,n}$ contains all surfaces which can be obtained by gluing surfaces of lower indices (the index of a surface being defined as $3g + n - 2$) through the plumbing fixture: they are interpreted as Feynman graphs containing propagators.
For the gluing to make sense, the local coordinates for surfaces in $\mc F_{g,n}$ are inherited from the ones on the lower-dimensional surfaces.
The remaining regions $\mc V_{g,n}$ (together with a choice of local coordinates, which must agree with the ones in $\mc F_{g,n}$ on common boundaries) define the fundamental $g$-loop $n$-point interaction.\footnotemark{}
\footnotetext{%
	Note that $\mc V_{g,n}$ and local coordinates are background independent~\cite{Sen:1994:ProofLocalBackground,Sen:1994:QuantumBackgroundIndependence,Sen:1996:BackgroundIndependentAlgebraic,Sen:2018:BackgroundIndependenceClosed}.
}%
These vertices then provide the interactions of the string field theory (SFT) action (see~\cite{Erbin:2021:StringFieldTheory,Zwiebach:1993:ClosedStringField,Erler:2020:FourLecturesClosed} for reviews).
In this paper, we will focus on classical vertices, i.e.~$g = 0$ (Riemann spheres).

This structure of the string vertices leads to two problems in SFT.
First, these vertices are not known explicitly except at the lowest orders.
Second, the action is non-polynomial.
In this paper, we focus on the first aspect.
While it is known in practice how to build vertices recursively once a 3-point $g = 0$ vertex is chosen, there is little chance to implement this in practice beyond the first few orders.\footnotemark{}
\footnotetext{%
	Note that different choices of vertices are related by field redefinitions~\cite{Zwiebach:1993:ClosedStringField}.
}%
For this reason, vertices which can be characterized by some geometric principle, such as minimal area~\cite{Zwiebach:1993:ClosedStringField} or hyperbolic~\cite{Moosavian:2020:HyperbolicGeometrySuperstring,Moosavian:2019:HyperbolicGeometryClosed-1,Moosavian:2019:HyperbolicGeometryClosed-2,Costello:2019:HyperbolicStringVertices} vertices, are more promising, in particular following the recent developments from~\cite{Firat:2023:BootstrappingClosedString,Firat:2023:HyperbolicStringTadpole}.
However, beyond the cubic order, these vertices have been characterized only numerically, and other local coordinates may be preferable for simple computations.
One particularly simple choice is $\mathrm{SL}(2, \C)$ vertices because they are amenable to analytic computations.
Usage examples of $\mathrm{SL}(2, \R)$ (open string) and $\mathrm{SL}(2, \C)$ include $D$-instantons and cosmological constant in open-closed SFT~\cite{Sen:2020:DinstantonPerturbationTheory,Sen:2021:DinstantonsStringField,Eniceicu:2022:ZZAnnulusOnepoint,Maccaferri:2022:ClassicalCosmologicalConstant}.

The $\mathrm{SL}(2, \C)$ cubic vertex was described in~\cite{Erler:2017:OneLoopTadpole} (see also~\cite{Zwiebach:1988:ConstraintsCovariantTheories,Sonoda:1989:HermiticityCPTString,Sonoda:1990:CovariantClosedString,Polchinski:2005:StringTheory-1}), as well as the 1-loop 1-point vertex (torus tadpole), though no explicit local coordinates in the vertex region were provided.
Lower-order vertices for open-closed SFT appeared in~\cite{Maccaferri:2022:ClassicalCosmologicalConstant}.
For minimal area vertices, the cubic vertex is a generalization of the open string Witten's vertex~\cite{Witten:1986:NoncommutativeGeometryString,Sonoda:1990:CovariantClosedString}, the quartic and quintic vertices were built numerically in~\cite{Moeller:2004:ClosedBosonicString,Moeller:2007:ClosedBosonicString-1,Moeller:2007:ClosedBosonicString-2} (a more general approach based on machine learning has been proposed recently in~\cite{Erbin:2022:Characterizing4stringContact} and implemented for the quartic vertex).
Finally, the cubic hyperbolic vertex has been described in~\cite{Firat:2021:HyperbolicThreestringVertex}, while the quartic and torus tadpole are discussed in~\cite{Firat:2023:BootstrappingClosedString,Firat:2023:HyperbolicStringTadpole}, with local coordinates given as a series expansion.

The main advantage of the $\mathrm{SL}(2, \C)$ vertex is that most computations can be performed analytically.
For example, we derive explicit formulas for the parametrization of the vertex region and its volume (to be compared with the minimal area quartic vertex, where the region is given by a fit or a neural network~\cite{Moeller:2004:ClosedBosonicString,Erbin:2022:Characterizing4stringContact}).
Finally, we provide the necessary and sufficient constraints on the local coordinates.

Another motivation for having an explicit quartic vertex is to study its decomposition using auxiliary fields in order to obtain a cubic formulation of closed SFT~\cite{Sonoda:1990:CovariantClosedString,Erbin:2018:CubicClosedString}.
Indeed, most successes achieved in open SFT -- in particular, analytic solutions and understanding tachyon condensation -- were possible thanks to the cubic nature of its interactions (see~\cite{Erler:2022:FourLecturesAnalytic} for a review).
Closed SFT is non-polynomial and solutions must be studied using both level and interaction truncations~\cite{Moeller:2004:ClosedBosonicString,Yang:2005:ClosedStringTachyon,Yang:2005:DilatonDeformationsClosed,Yang:2005:DilatonDeformationsClosed,Yang:2005:TestingClosedString,Moeller:2007:NonperturbativeClosedString,Moeller:2007:ClosedBosonicString-1,Moeller:2007:ClosedBosonicString-2,Moeller:2008:TachyonLumpClosed,Scheinpflug:2023:ClosedStringTachyon}.
Hence, important issues such as the closed string tachyon vacuum and the nature of classical solutions are still not understood.
The Hubbard--Stratonovich transformation~\cite{Stratonovich:1957:MethodCalculatingQuantum,Hubbard:1959:CalculationPartitionFunctions} introduces auxiliary fields to decompose higher-order interactions into cubic interactions.
This idea has recently been implemented for open string with stubs~\cite{Erbin:2023:OpenStringStub}, which is non-polynomial but can be made cubic with a single auxiliary field.

\paragraph{Outline}

In \Cref{sec:setup}, we recall how the SFT action is built, then we describe the $\mathrm{SL}(2, \C)$ cubic vertex from~\cite{Erler:2017:OneLoopTadpole}.
Next, in \Cref{sec:S04-propagator}, we obtain the regions of the moduli space corresponding to Feynman diagrams with propagators and provide the corresponding local coordinates.
Finally, in \Cref{sec:vertex}, we discuss how to build local coordinates in a fundamental domain of the vertex region.

\section{Setup}
\label{sec:setup}

\subsection{String field theory vertices}
\label{sec:setup:vertex}

At tree-level, the Riemann surfaces are $n$-punctured spheres.
The moduli space is parametrized by $n - 3$ complex numbers, $\mc M_{0,n} \sim \C^{n-3}$, since three punctures can be fixed using a $\group{SL}(2, \C)$ transformation.
The global coordinate on the sphere is denoted by $z$, the punctures are located at $z = z_i$ ($i = 1, \ldots, n$), and the three fixed punctures are taken to be located at $(z_1, z_2, z_3) = (0, 1, \infty)$.

The local coordinates $w_i$ are related to the global coordinates by holomorphic maps $f_i(w)$
\begin{equation}
	z = f_i(w_i),
	\qquad
	z_i = f_i(0)
\end{equation}
which satisfy the following conditions~\cite{Zwiebach:1993:ClosedStringField,Sen:2015:OffshellAmplitudesSuperstring,Erler:2020:FourLecturesClosed,Erbin:2021:StringFieldTheory}:
\begin{enumerate}
	\item The local coordinates lie in the unit disk:
	\begin{equation}
		\abs{w_i} \le 1.
	\end{equation}

	\item The local coordinates are defined up to rotations by global phases:
	\begin{equation}
		f_i(w) \sim f_i\big(\e^{\I \theta_i} w\big).
	\end{equation}

	\item The local coordinates transform under each other under permutations $\sigma \in S_n$ of the punctures:
	\begin{equation}
		z_{\sigma(i)} = \sigma \circ z_i
		\quad \Longrightarrow \quad
		\sigma \circ f_i(w) = f_{\sigma(i)}\big(\e^{\I \theta_{\sigma,i}} w\big),
	\end{equation}
	where the phases $\theta_{\sigma,i}$ are allowed by the previous point.
\end{enumerate}
Given $n$ string states $A_i$ in the CFT Hilbert space, the $n$-point interaction in the action is given by computing the CFT correlation function of the $n$ states located at the origin ($w_i = 0$) of the local coordinate patches, mapped back to the surface $\Sigma_{0,n}$ through the maps $f_i$, and integrating the moduli $z_i$ over the vertex region $\mc V_{0,n}$:
\begin{equation}
	\mc V_{0,n}(A_1, \ldots, A_n)
		:= \int_{\mc V_{0,n}} \prod_{i=4}^n \dd^2 z_i \,
			\Braket{\text{ghosts} \times \prod_{i=1}^n f_i \circ A_i(0)}_{\Sigma_{0,n}}.
\end{equation}
We do not write explicitly the ghost contributions since they are not relevant for this paper.

Two surfaces $\Sigma_{0,n_1}^{(1)}$ and $\Sigma_{0,n_2}^{(2)}$ with local coordinates
\begin{equation}
	z^{(1)}
		= f_i^{(1)}\big(w_i^{(1)}\big),
	\qquad
	z^{(2)}
		= f_j^{(2)}\big(w_i^{(2)}\big)
\end{equation}
can be glued together using the plumbing fixture relation:
\begin{equation}
	\label{eq:plumbing}
	w_{n_1}^{(1)} w_{n_2}^{(2)}
		= q,
	\qquad
	\abs{q} \le 1.
\end{equation}
Gluing two surfaces in this way is conformally equivalent to connecting the surfaces by a cylinder.
In terms of the global coordinates, the plumbing fixture becomes:
\begin{equation}
	z^{(1)}
		= f_{n_1}^{(1)}\left( \frac{q}{f_{n_2}^{(2),-1}(z^{(2)})} \right),
\end{equation}
where $f_n^{(2),-1}(z)$ is the inverse of $f_n^{(2)}$.
We get a new surface $\Sigma_{0, n_1 + n_2 - 2}$ with global coordinate $Z = z^{(1)}$ (convention) and local coordinates $F_I(W_I)$ induced from the coordinates on $\Sigma_{0,n_1}^{(1)}$ and $\Sigma_{0,n_2}^{(2)}$.
For the punctures on $\Sigma_{0,n_2}^{(1)}$, we have
\begin{equation}
	F_i(W_i)
		= f_i^{(1)}(W_i),
	\qquad
	W_i
		= w_i^{(1)},
	\qquad
	i = 1, \ldots, n_1 - 1,
\end{equation}
while, for the punctures on $\Sigma_{0,n}^{(1)}$, we need to convert $z^{(2)}$ using the plumbing fixture first:
\begin{equation}
	\begin{gathered}
	F_{n_1 - 1 + j}(W_{n_1 - 1 + j})
		= f_{n_1}^{(1)}\left( \frac{q}{f_{n_2}^{(2),-1}(f_j^{(2)}\big( W_{n_1 - 1 + j}) \big)} \right),
	\\
	W_{n_1 - 1 + j}
		= w_j^{(2)},
	\qquad
	j = 1, \ldots, n_2 - 1,
	\end{gathered}
\end{equation}
The moduli parameters $Z_I$ of the new surface can be read as $F_I(0) = Z_I$.
However, the punctures $(Z_1, Z_2, Z_3)$ are generically not located at $(0, 1, \infty)$ and we must perform a $\mathrm{SL}(2, \C)$ transformation to map them back before determining the moduli.
The latter then depend on $q$ and varying it generates a family of surfaces which cover some subspace $\mc F_{0,n_1 + n_2 - 2}^{(1)}$ of the moduli space.
Other surfaces can be obtained by permuting which coordinates $F_I$ are induced from coordinates on one surface or the other.

It is convenient to parametrize $q$ as
\begin{equation}
	\label{eq:q-stheta}
	q
		= \e^{- s + \I \theta},
	\qquad
	s \in \R_+,
	\quad
	\theta \in [0, 2\pi]
\end{equation}
as it will make the geometrical picture clearer.

For $\group{SL}(2, \C)$ vertices, the maps $f_i$ are given by elements of the group $\group{GL}(2, \C)$ since it is simpler to not impose that the determinant is one.
The goal of this paper is to compute the closed string vertex $\mc V_{0,4}$.

\subsection{Cubic vertex}
\label{sec:setup:S03}

The $3$-point vertex matches the scattering amplitude since $\dim \mc M_{0,3} = 0$.
The three punctures are fixed at
\begin{equation}
	z_0
		= 0,
	\qquad
	z_1
		= 1,
	\qquad
	z_\infty
		= \infty.
\end{equation}
If ones does not consider generalized sections~\cite{Sen:2015:OffshellAmplitudesSuperstring}, there is a unique family of $\group{SL}(2, \C)$ vertices parametrized by a single parameter~\cite{Erler:2017:OneLoopTadpole}.
The maps are denoted by $\{ f_0, f_1, f_\infty \}$.

We look for functions $f_i \in \group{SL}(2, \C)$ such that
\begin{equation}
	f_0(0)
		= 0,
	\qquad
	f_1(0)
		= 1,
	\qquad
	f_\infty(0)
		= \infty,
\end{equation}
which transform into each other under the permutation group $S_3 = \Span \{ g_{01}, g_{0\infty} \}$ (see \Cref{app:sl2c})
\begin{equation}
	g_{\sigma} \circ f_i(w) = f_{\sigma(i)}\big(\e^{\I \theta_{\sigma,i}} w\big),
\end{equation}
where we allow for phases $\theta_{\sigma,i}$.

The condition $f_0(0) = 0$ tells that
\begin{equation}
	f_0(w)
		= \frac{w}{\alpha w + \beta}.
\end{equation}
Imposing the constraint that it is invariant under $g_{1 \infty}$
\begin{equation}
	g_{1 \infty} \circ f_0(w)
		= f_0(\e^{\I \theta} w),
\end{equation}
we get the conditions:
\begin{equation}
	\e^{\I\theta}
		= - 1,
	\qquad
	\alpha
		= \frac{1}{2}.
\end{equation}
The functions $f_1$ and $f_\infty$ are obtained by applying $g_{01}$ and $g_{0\infty}$ to $f_0$, such that
\begin{equation}
	\label{eq:S03:fw}
	\boxed{
	f_0(w) = \frac{2 w}{2\beta + w},
	\qquad
	f_1(w) = \frac{2\beta - w}{2\beta + w},
	\qquad
	f_\infty(w) = \frac{2\beta + w}{2w}.
	}
\end{equation}
Note that the determinant is not one because normalizating leads to more complicated expressions; as a consequence, we have to remember that equality between two maps holds up to a scale factor.

We can also solve for $w$ in terms of $f_i(w)$
\begin{equation}
	\label{eq:3pt:w-from-f}
	\boxed{
	w_0
		= 2\beta \, \frac{f_0(w)}{2 - f_0(w)},
	\qquad
	w_1
		= 2\beta \, \frac{1 - f_1(w)}{1 + f_1(w)},
	\qquad
	w_\infty
		= 2\beta \, \frac{1}{2 f_\infty(w) - 1}.
	}
\end{equation}
This shows that $\beta \in \R$ since any phase of $\beta$ can be absorbed in the definition of the local coordinates $w_i$.
Moreover, this also means that $\beta$ is related to the stub parameter of the theory (see~\cite{Erbin:2021:StringFieldTheory,Chiaffrino:2021:QFTStubs,Erbin:2023:OpenStringStub,Schnabl:2023:OpenStringField,Scheinpflug:2023:ClosedStringTachyon} for definitions and recent discussions).
The stub parameter determines which parts of the Riemann surfaces (or Feynman diagrams) is part of the interaction or part of the propagator, and it provides a natural UV cutoff.

\begin{figure}[htp]
	\centering
	\subcaptionbox{$\beta = 1.5$}{%
		\includegraphics[scale=0.8]{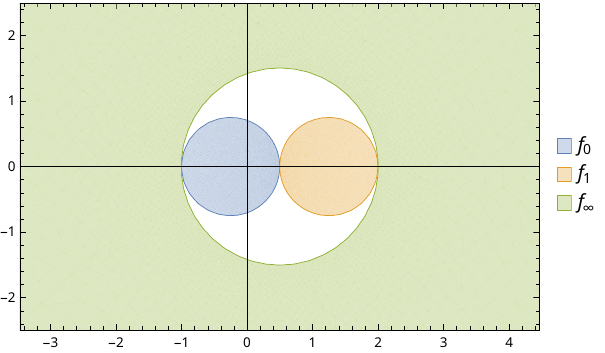}
	}
	\medskip

	\subcaptionbox{$\beta = 2$}{%
		\includegraphics[scale=0.8]{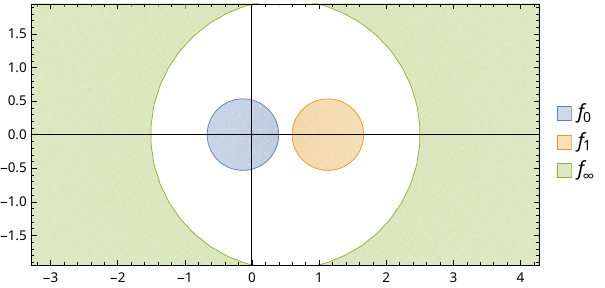}
	}
	\caption{%
		Local coordinates patches in the plane $z \in \C$ for the $3$-point vertex.
	}
	\label{fig:S03:coord}
\end{figure}

Coordinate patches should not overlap~\cite{Erler:2020:FourLecturesClosed}, which leads to the following bound on the parameter $\beta$:
\begin{equation}
	\label{eq:S03:bound-beta}
	\beta
		\ge \beta_{0,3}
		:= \frac{3}{2}.
\end{equation}
This inequality can be solved by parametizing $\beta$ as
\begin{equation}
	\label{eq:beta-s0}
	\beta
		= \frac{3}{2} \, \e^{\frac{s_0}{2}}
\end{equation}
where $s_0$ is the stub parameter.
Indeed, adding a stub of length $s_0$ is equivalent to multiplying the interaction vertices by $\e^{- s_0 L_0^+}$, which is equivalent to a rescaling of the local coordinates by $\e^{- s_0}$.
The local coordinate patches for $\beta \in \{ 1.5, 2 \}$ are shown in \Cref{fig:S03:coord}.

\section{Quartic propagator graphs}
\label{sec:S04-propagator}

A $4$-punctured sphere can be obtained by gluing two $3$-point vertices with a propagator.
The number of different gluing is given by the number of partitions of the set $\{ 1, 2, 3, 4 \}$ in two (unordered sets): hence, there are three possibilities~\cite{Sonoda:1990:CovariantClosedString}, the $s$-, $t$- and $u$-channels (respectively, $12 \to 34$, $13 \to 24$ and $14 \to 23$), as expected from usual QFT.
To understand this, we number all punctures, except the ones to be glued together, on each surface, and we consider the gluing of all inequivalent arrangements of numbers (up to permutations on the same surface).

We introduce the following notations for the moduli space of $4$-punctured spheres:
\begin{equation}
	\mc M_{0,4}
		= \mc V_{0,4} \cup \mc F_{0,4},
	\qquad
	\mc F_{0,4}
		= \mc F_{0,4}^{(s)} \cup \mc F_{0,4}^{(t)} \cup \mc F_{0,4}^{(u)},
\end{equation}
where $\mc V_{0,4}$ is the fundamental region not covered by any graphs with propagators.
We denote quantities related to 4-punctured spheres with capital letters.
Hence, the global and local coordinates on the sphere are denoted by $Z$ and $W_a$ ($a = 1, \ldots, 4$), the latter being associated to maps $F_a$ such that
\begin{equation}
	Z
		= F_a(W_a),
	\qquad
	Z_a
		= F_a(0).
\end{equation}
The puncture $Z_4$ is taken to cover the moduli space $\mc M_{0,4}$ and the three others are fixed
\begin{equation}
	Z_1
		= 0,
	\qquad
	Z_2
		= 1,
	\qquad
	Z_3
		= \infty,
	\qquad
	Z_4 \in \C.
\end{equation}

\subsection{Fundamental domain}
\label{sec:S04:fundamental-domain}

This is helpful for two reasons~\cite{Moeller:2004:ClosedBosonicString}.
First, we can deduce constraints on the maps:
\begin{enumerate}
	\item The fundamental vertex must be invariant under permutations of the four punctures (and local coordinate maps).

	\item The regions covered by the $s$-, $t$- and $u$-channels must mapped to each other under permutations.
\end{enumerate}
The second point implies that it is sufficient to study the gluing in the $s$-channel to deduce the local coordinates and Feynman regions of the $t$- and $u$-channels.
Second, we can identify a fundamental domain (FD) for the group $S_4$ and restrict all the computations to that domain, before finding the maps elsewhere by applying elements of $S_4$.

Permuting the local coordinates at one point $Z_4$ of the moduli space maps to another point $Z_4'$.
Indeed, permutations are implemented by $\group{SL}(2, \C)$ transformations of the global coordinates, which keep fix only three points: hence, the fourth point corresponding to the moduli is mapped to another point.
\begin{equation}
	\label{eq:map-coord-S4-gen}
	S_{14}(Z)
		:= \frac{Z_4 - Z}{Z_4 - 1},
	\qquad
	S_{24}(Z)
		:= \frac{Z}{Z_4},
	\qquad
	S_{34}(Z)
		:= \frac{Z (1 - Z_4)}{Z - Z_4},
\end{equation}
see \Cref{app:sl2c} for more information on $\group{SL}(2, \C)$ and permutations.
These maps acting on the global coordinate (or local coordinate maps) induce maps on the moduli space, which are found by evaluating the previous functions at the locations of the previous fixed punctures:
\begin{equation}
	\label{eq:map-Z4-S4-gen}
	\begin{gathered}
	S_{14}(0)
		= g_{1\infty}(Z_4),
	\qquad
	S_{24}(1)
		= g_{0\infty}(Z_4),
	\\
	S_{34}(\infty)
		= g_{01}(Z_4),
	\end{gathered}
\end{equation}
where the RHS are defined in \eqref{eq:sl2c:S3-perm}.
On the moduli space, this reduces to the action of the $S_3$ permutation group.
Other permutations of $S_4$ induce the same maps on the moduli space.
We will be interested in particular in:
\begin{equation}
	\label{eq:map-coord-S4-S3}
	S_{12}(Z)
		:= g_{0\infty}(Z).
	\qquad
	S_{13}(Z)
		:= g_{01}(Z),
	\qquad
	S_{23}(Z)
		:= g_{1\infty}(Z),
\end{equation}
The new moduli locations are found by evaluating the maps at $Z = Z_4$, and one recovers the same maps as in \eqref{eq:map-Z4-S4-gen}.
The only subtlety is that the ordering of the punctures is different after the maps \eqref{eq:map-coord-S4-gen} and \eqref{eq:map-coord-S4-S3}.

In order to find a fundamental domain, we can plot the image of the unit circle, parametrized by the angle $\nu \in [0, 2\pi)$, centered at the origin under each of the maps: the FD must be a domain where no region overlaps.
The notation is chosen to avoid confusion with the twist parameter $\theta$ of the plumbing fixture and because $\nu$ will be identified with the true anomaly \eqref{eq:map-angles-ellipse} of the ellipse defining the boundary of the $s$-channel region.
The images are: the unit circles centered at the origin and at $1$, and the infinite line $\Re Z_4 = 1/2$:
\begin{equation}
	g_{01}(\e^{\I \nu})
		= 1 - \e^{\I \nu},
	\qquad
	g_{0\infty}(\e^{\I \nu})
		= \e^{- \I \nu},
	\qquad
	g_{1\infty}(\e^{\I \nu})
		= \frac{1}{2} - \frac{\I}{2} \, \frac{\sin \nu}{1 - \cos \nu}.
\end{equation}
The three regions meet at two points $Q$ and $\bar Q$
\begin{equation}
	Q
		:= \e^{\I \pi / 3}
		= \frac{1}{2} + \frac{\I \sqrt{3}}{2}.
\end{equation}
We also add complex conjugation to the set of maps because we want obtain real quantities.

\begin{figure}[ht]
	\centering
	\includegraphics[scale=1.5]{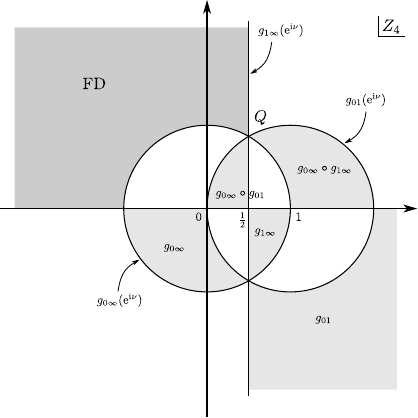}
	\caption{%
		Fundamental domain of $S_4$ (FD, in grey) and its images under the different $S_3$ maps (lighter grey).
		The images of the unit circle are also displayed.
	}
	\label{fig:FD-S4}
\end{figure}

We choose to work with the following FD (\Cref{fig:FD-S4}, following the convention from~\cite{Moeller:2004:ClosedBosonicString}):
\begin{equation}
	\label{eq:fd-cond-C}
	\Re Z_4 \le \frac{1}{2},
	\qquad
	\abs{Z_4} \ge 1,
	\qquad
	\Im Z_4 \ge 0.
\end{equation}
The images of the FD under the maps \eqref{eq:map-Z4-S4-gen} and under the additional maps:
\begin{equation}
	g_{0\infty} \circ g_{01}(Z_4)
		= \frac{Z_4 - 1}{Z_4},
	\qquad
	g_{0\infty} \circ g_{1\infty}
		= \frac{1}{1 - Z_4}
\end{equation}
are displayed on \Cref{fig:FD-S4}.
Due to the equalities
\begin{equation}
	\begin{gathered}
	g_{01} \circ g_{0\infty}
		= g_{0\infty} \circ g_{1\infty},
	\qquad
	g_{01} \circ g_{1\infty}
		= g_{0\infty} \circ g_{01},
	\\
	g_{1\infty} \circ g_{01}
		= g_{0\infty} \circ g_{1\infty},
	\qquad
	g_{1\infty} \circ g_{0\infty}
		= g_{0\infty} \circ g_{01},
	\end{gathered}
\end{equation}
there are no more images than the five displayed (plus the complex conjugated domains).
Hence, the complex plane is separated in $12$ regions under $S_4$ and complex conjugation.
Note that the images under a single map lie in the lower half-plane.

\subsection{Local coordinates}

Following \Cref{sec:setup:vertex}, we glue two 3-punctured spheres $\Sigma_{0,3}^{(1)}$ and $\Sigma_{0,3}^{(2)}$ to obtain a 4-punctured sphere $\Sigma_{0,4}$.
Coordinates and maps related to the 3-punctured spheres are written with lower-case letters, with the sphere number given as a superscript:
\begin{equation}
	z^{(1)}
		= f^{(1)}_i\big(w_i^{(1)} \big),
	\qquad
	z^{(2)}
		= f^{(2)}_i\big(w_i^{(2)} \big),
\end{equation}
where $i = 0, 1, \infty$.
By convention, we will glue together the punctures located at $z = \infty$:
\begin{equation}
	\label{eq:S04-plumbing}
	w_\infty^{(1)} w_\infty^{(2)}
		= q,
	\qquad
	\abs{q} \le 1,
\end{equation}
such that
\begin{equation}
	\label{eq:S04-plumbing-z}
	z^{(1)}
		= \frac{1}{2} \left( 1 + \frac{4 \beta^2}{q} \, \frac{1}{2 z^{(2)} - 1} \right),
\end{equation}
by using \eqref{eq:3pt:w-from-f} in \eqref{eq:S04-plumbing}.
In order to find the maps $F_a(W_a)$ we need to relate them to $Z$, which can itself be related to $z^{(1)}$ and $z^{(2)}$, themselves related to the $f_i^{(1)}$ and $f_i^{(2)}$.
We identify the global coordinate of $\Sigma_{0,4}$ with the one of $\Sigma_{0,3}^{(1)}$
\begin{equation}
	Z = z^{(1)}.
\end{equation}

\paragraph{$s$-channel}

\begin{figure}[ht]
	\centering
	\includegraphics[scale=0.8]{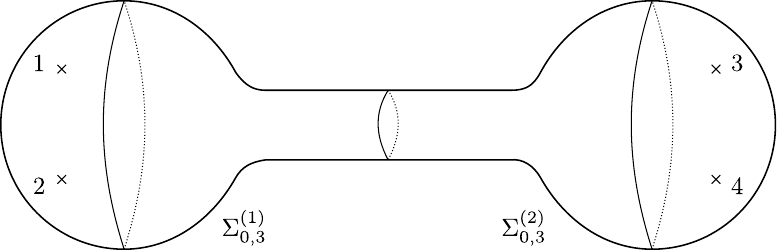}
	\caption{$s$-channel graph with $12 \to 34$.}
	\label{fig:simple:4pt-s}
\end{figure}

To obtain the $s$-channel graph (\Cref{fig:simple:4pt-s}), we need to perform the following identifications:
\begin{equation}
	W_1
		= w_0^{(1)},
	\qquad
	W_2
		= w_1^{(1)},
	\qquad
	W_3
		= w_0^{(2)},
	\qquad
	W_4
		= w_1^{(2)}.
\end{equation}

The first transition function is found through
\begin{equation}
	F_1^{(s)} \big( w_0^{(1)} \big)
		= Z
		= z^{(1)}
		= f_0^{(1)}\big( w_0^{(1)} \big).
\end{equation}
Similarly, for the second function one finds
\begin{equation}
	F_2^{(s)} \big( w_1^{(1)} \big)
		= Z
		= z^{(1)}
		= f_1^{(1)}\big( w_1^{(1)} \big).
\end{equation}

In order to find the two others, we need to relate $z^{(1)}$ to $z^{(2)}$ using \eqref{eq:S04-plumbing-z}.
For the third function, we find
\begin{align*}
	F_3^{(s)} \big(w_0^{(2)}\big)
		&
		= Z
		= z^{(1)}
		= \frac{1}{2} \left( 1 + \frac{4 \beta^2}{q} \, \frac{1}{2 z^{(2)} - 1} \right)
		= \frac{1}{2} \left( 1 + \frac{4 \beta^2}{q} \, \frac{1}{2 f_0^{(2)}(w_0^{(2)}) - 1} \right).
\end{align*}
Finally for the fourth function, one has
\begin{align*}
	F_4^{(s)} \big(w_1^{(2)}\big)
		&
		= Z
		= z^{(1)}
		= \frac{1}{2} \left( 1 + \frac{4 \beta^2}{q} \, \frac{1}{2 z^{(2)} - 1} \right)
		= \frac{1}{2} \left( 1 + \frac{4 \beta^2}{q} \, \frac{1}{2 f_1^{(2)}(w_1^{(2)}) - 1} \right).
\end{align*}

After simpliciations, the four local coordinate maps are
\begin{equation}
	\begin{aligned}
	F_1^{(s)}(W)
		&
		= \frac{2 W}{2\beta + W},
	\\
	F_2^{(s)}(W)
		&
		= \frac{2\beta - W}{2\beta + W},
	\\
	F_3^{(s)}(W)
		&
		= \frac{1}{2} \left( 1 + \frac{4 \beta^2}{q} \,
		\frac{2\beta + W}{3 W - 2\beta} \right),
	\\
	F_4^{(s)}(W)
		&
		= \frac{1}{2} \left( 1 - \frac{4 \beta^2}{q} \,
		\frac{2\beta + W}{3 W - 2\beta} \right).
	\end{aligned}
\end{equation}
The position of the punctures can be found by evaluating the maps at $W = 0$:
\begin{equation}
	\label{eq:punc-s}
	F_1^{(s)}(0)
		= 0,
	\quad
	F_2^{(s)}(0)
		= 1,
	\quad
	F_3^{(s)}(0)
		= \frac{1}{2} \left( 1 - \frac{4 \beta^2}{q} \right),
	\quad
	F_4^{(s)}(0)
		= \frac{1}{2} \left( 1 + \frac{4 \beta^2}{q} \right).
\end{equation}

In order to find which region of the moduli space is covered we need to perform a transformation $g_s \in \group{SL}(2, \C)$ to move $F_3(0) \to Z_3 = \infty$ while leaving $F_1(0)$ and $F_2(0)$ fixed:
\begin{equation}
	g_s\big(F_1^{(s)}(0) \big)
		= F_1^{(s)}(0),
	\qquad
	g_s\big(F_2^{(s)}(0) \big)
		= F_2^{(s)}(0),
	\qquad
	g_s\big(F_3^{(s)}(0) \big)
		= \infty,
\end{equation}
after which the value of the moduli parameter can be read from $Z_4 = g_s(F_4(0))$.
The solution is:
\begin{equation}
	g_s(z)
		= \frac{\Big( F_3^{(s)}(0) - 1 \Big) \, z}{F_3^{(s)}(0) - z}.
\end{equation}
The moduli parameter in the channel reads then
\begin{equation}
	\label{eq:Z4-s-q}
	\boxed{
	Z_4^{(s)}(q)
		:= g_s \big(F_4^{(s)}(0) \big)
		= \frac{q}{16 \beta^2} \left( 1 + \frac{4 \beta^2}{q} \right)^2.
	}
\end{equation}
As expected, the region covered by $Z_4$ includes $Z_3$:
\begin{equation}
	\lim_{\abs{q} \to 0} Z_4^{(s)}
		= Z_3
		= \infty.
\end{equation}
We can invert the relation to express $q$ in terms of $Z_4$:
\begin{equation}
	q
		= 4 \beta^2 \left[ (2 Z_4 - 1) - 2 \sqrt{Z_4 (Z_4 - 1)} \right].
\end{equation}
The solution is quadratic and has two solutions: only one satisfies $|q| \le 1$ when $Z_4$ is in the $s$-channel region.

We now compute the functions $F_a$ after performing the transformation $g_s$.
Note that even if $F_1(0)$ and $F_2(0)$ are invariant, this is not the case of the complete function.
We find
\begin{equation}
	\label{eq:4pt-locmaps-s}
	\boxed{
	\begin{aligned}
	\tilde F_1^{(s)}(W)
		&
		= (4 \beta^2 + q) \;
			\frac{2 W}{W (3 q + 4 \beta^2) + 2\beta (4 \beta^2 - q)},
	\\
	\tilde F_2^{(s)}(W)
		&
		= (4 \beta^2 + q) \;
			\frac{W - 2\beta}{W (3 q - 4 \beta^2) - 2\beta (4 \beta^2 + q)},
	\\
	\tilde F_3^{(s)}(W)
		&
		= \frac{4 \beta^2 + q}{16 \beta^2 q} \;
			\frac{2\beta (4 \beta^2 - q) + W (3 q + 4 \beta^2)}{2 W},
	\\
	\tilde F_4^{(s)}(W)
		&
		= \frac{4 \beta^2 + q}{16 \beta^2 q} \;
			\frac{2\beta (4 \beta^2 + q) - W (3 q - 4 \beta^2)}{W - 2\beta}.
	\end{aligned}
	}
\end{equation}
where we defined $\tilde F_i^{(s)} := g_s \circ F_i^{(s)}$.

\paragraph{$t$-channel}

To obtain the $t$-channel graph, we need to perform the following identification of the local coordinates:
\begin{equation}
	W_1
		= w_0^{(1)},
	\qquad
	W_2
		= w_0^{(2)},
	\qquad
	W_3
		= w_1^{(1)},
	\qquad
	W_4
		= w_1^{(2)}.
\end{equation}
The computations are the same as in the previous case, but the results can also be obtained by permuting the second and third punctures.
The transitions functions are:
\begin{equation}
	\begin{aligned}
		F_1^{(t)}(W)
			&
			= \frac{2 W}{2\beta + W},
		\\
		F_2^{(t)}(W)
			&
			= \frac{1}{2} \left( 1 + \frac{4 \beta^2}{q} \,
				\frac{2\beta + W}{3 W - 2\beta} \right),
		\\
		F_3^{(t)}(W)
			&
			= \frac{2\beta - W}{2\beta + W},
		\\
		F_4^{(t)}(W)
			&
			= \frac{1}{2} \left( 1 - \frac{4 \beta^2}{q} \,
				\frac{2\beta + W}{3 W - 2\beta} \right),
	\end{aligned}
\end{equation}
from which one reads the positions of the punctures:
\begin{equation}
	\label{eq:punc-t}
	F_1^{(t)}(0)
		= 0,
	\quad
	F_2^{(t)}(0)
		= \frac{1}{2} \left( 1 - \frac{4 \beta^2}{q} \right),
	\quad
	F_3^{(t)}(0)
		= 1,
	\quad
	F_4^{(t)}(0)
		= \frac{1}{2} \left( 1 + \frac{4 \beta^2}{q} \right).
\end{equation}
The transformation
\begin{equation}
	g_t(z)
		= \frac{1 - F_2^{(t)}(0)}{F_2^{(t)}(0)} \, \frac{z}{1 - z}
\end{equation}
moves $F_2^{(t)}(0) \to 1$ and $F_3^{(t)}(0) \to \infty$.
This leads to the following expression for the $t$-channel moduli:
\begin{equation}
	\label{eq:punc-t-ord}
	\boxed{
	Z_4^{(t)}(q)
		= \left( \frac{q + 4 \beta^2}{q - 4 \beta^2} \right)^2.
	}
\end{equation}
As expected, the region covered by $Z_4$ includes $Z_2$
\begin{equation}
	\lim_{\abs{q} \to 0} Z_4^{(t)}
		= Z_2
		= 1.
\end{equation}
We can invert the relation to express $q$ in terms of $Z_4$:
\begin{equation}
	q
		= 4 \beta^2 \, \frac{\sqrt{Z_4} - 1}{\sqrt{Z_4} + 1}.
\end{equation}

The local coordinates after reordering the punctures are:
\begin{equation}
	\label{eq:4pt-locmaps-t}
	\boxed{
	\begin{aligned}
	\tilde F_1^{(t)}(W)
		&
		= \frac{4 \beta^2 + q}{4 \beta^2 - q} \;
			\frac{2 W}{W - 2\beta},
	\\
	\tilde F_2^{(t)}(W)
		&
		= - \frac{4 \beta^2 + q}{4 \beta^2 - q} \;
			\frac{W (3 q + 4 \beta^2) + 2\beta (4 \beta^2 - q)}{W (3 q - 4 \beta^2) - 2\beta (4 \beta^2 + q)},
	\\
	\tilde F_3^{(t)}(W)
		&
		= \frac{4 \beta^2 + q}{4 \beta^2 - q} \;
			\frac{W - 2\beta}{2 W},
	\\
	\tilde F_4^{(t)}(W)
		&
		= - \frac{4 \beta^2 + q}{4 \beta^2 - q} \;
			\frac{W (3 q - 4 \beta^2) - 2\beta (4 \beta^2 + q)}{W (3 q + 4 \beta^2) + 2\beta (4 \beta^2 - q)}.
	\end{aligned}
	}
\end{equation}

\paragraph{$u$-channel}

Finally, the following identifications give the $u$-channel graph:
\begin{equation}
	W_1
		= w_0^{(1)},
	\qquad
	W_2
		= w_0^{(2)},
	\qquad
	W_3
		= w_1^{(2)},
	\qquad
	W_4
		= w_1^{(1)}.
\end{equation}
The transitions functions are:
\begin{equation}
	\begin{aligned}
	F_1^{(u)}(W)
		&
		= \frac{2 W}{2\beta + W},
	\\
	F_2^{(u)}(W)
		&
		= \frac{1}{2} \left( 1 + \frac{4 \beta^2}{q} \,
		\frac{2\beta + W}{3 W - 2\beta} \right),
	\\
	F_3^{(u)}(W)
		&
		= \frac{1}{2} \left( 1 - \frac{4 \beta^2}{q} \,
		\frac{2\beta + W}{3 W - 2\beta} \right),
	\\
	F_4^{(u)}(W)
		&
		= \frac{2\beta - W}{2\beta + W},
	\end{aligned}
\end{equation}
from which we get:
\begin{equation}
	\label{eq:punc-u}
	\begin{gathered}
	F_1^{(u)}(0)
		= 0,
	\qquad
	F_2^{(u)}(0)
		= \frac{1}{2} \left( 1 - \frac{4 \beta^2}{q} \right),
	\\
	F_3^{(u)}(0)
		= \frac{1}{2} \left( 1 + \frac{4 \beta^2}{q} \right),
	\qquad
	F_4^{(u)}(0)
		= 1.
	\end{gathered}
\end{equation}

One can move move $F_2^{(u)}(0) \to 1$ and $F_3^{(u)}(0) \to \infty$ using the transformation
\begin{equation}
	g_u(z)
		= \frac{F_3^{(u)}(0) - F_2^{(u)}(0)}{F_2^{(u)}(0)} \, \frac{z}{F_3^{(u)}(0) - z}.
\end{equation}
This leads to the expression for $Z_4$ in the $u$-channel:
\begin{equation}
	\label{eq:punc-u-ord}
	\boxed{
	Z_4^{(u)}(q)
		= - \frac{16 \beta^2 q}{(q - 4 \beta^2)^2}.
	}
\end{equation}
As expected, the region covered by $Z_4$ includes $Z_1$
\begin{equation}
	\lim_{q \to 0} Z_4^{(u)}
		= Z_1
		= 0.
\end{equation}
We can invert the relation to express $q$ in terms of $Z_4$:
\begin{equation}
	q
		= \frac{4 \beta^2}{Z_4} \,
			\left[ Z_4 - 2 + 2 \sqrt{1 - Z_4} \right].
\end{equation}

The local coordinates after reordering the punctures are:
\begin{equation}
	\label{eq:4pt-locmaps-u}
	\boxed{
	\begin{aligned}
	\tilde F_1^{(u)}(W)
		&
		= \frac{16 \beta^2 q}{4 \beta^2 - q} \;
			\frac{2 W}{W (3 q - 4 \beta^2) - 2\beta (4 \beta^2 + q)},
	\\
	\tilde F_2^{(u)}(W)
		&
		= - \frac{1}{4 \beta^2 - q} \;
			\frac{W (3 q + 4 \beta^2) + 2\beta (4 \beta^2 - q)}{W - 2\beta},
	\\
	\tilde F_3^{(u)}(W)
		&
		= \frac{1}{4 \beta^2 - q} \;
			\frac{W (3 q - 4 \beta^2) - 2\beta (4 \beta^2 + q)}{2 W},
	\\
	\tilde F_4^{(u)}(W)
		&
		= \frac{16 \beta^2 q}{4 \beta^2 - q} \;
			\frac{W - 2 \beta}{W (3 q + 4 \beta^2) + 2\beta (4 \beta^2 - q)}.
	\end{aligned}
	}
\end{equation}

\subsection{Feynman regions}

\begin{figure}[htp]
	\centering
	\includegraphics[scale=0.6]{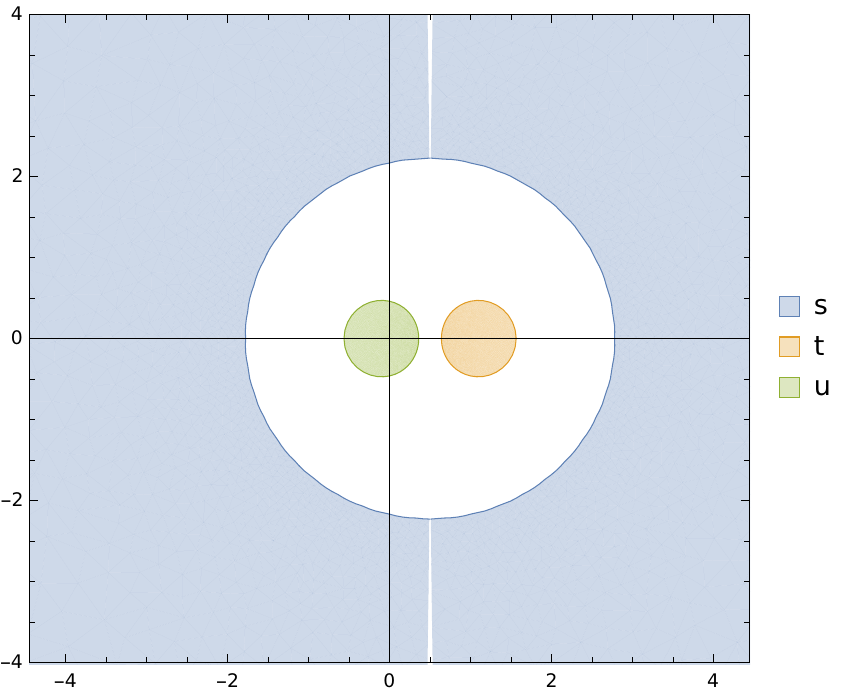}
	\caption{Feynman regions in moduli space for $\beta = \beta_{0,3} = 1.5$.}
	\label{fig:simple:M04-prop}
\end{figure}

In this subsection, we describe some additional properties of the Feynman regions.
We recall the formulas for the locations of the moduli parameter $Z_4$ in each channel:
\begin{subequations}
\label{eq:M04-stu-q}
\begin{align}
	Z_4^{(s)}(q)
		&
		= \frac{q}{16 \beta^2} \left( 1 + \frac{4 \beta^2}{q} \right)^2
		= \cosh^2 H(q),
	\\
	Z_4^{(t)}(q)
		&
		= \left( \frac{q + 4 \beta^2}{q - 4 \beta^2} \right)^2
		= \frac{\cosh^2 H(q)}{\sinh^2 H(q)},
	\\
	Z_4^{(u)}(q)
		&
		= - \frac{16 \beta^2 q}{(q - 4 \beta^2)^2}
		= - \frac{1}{\sinh^2 H(q)},
\end{align}
\end{subequations}
where
\begin{equation}
	\label{eq:Z4-stu-H}
	H(q) = \ln \frac{2 \beta}{\sqrt{q}}.
\end{equation}
The regions covered by varying $\abs{q} \le 1$ for each channel are given in \Cref{fig:simple:M04-prop}.

We can also rewrite these formulas using $q = \e^{- s + \I \theta}$, see \eqref{eq:q-stheta}:
\begin{subequations}
\label{eq:summary-punc}
\begin{align}
	\label{eq:summary-punc-s}
	Z_4^{(s)}(s, \theta)
		&
		= \frac{1}{2}
			+ \left( \beta^2 + \frac{1}{16 \beta^2} \right)
				\cosh(s - \I \theta)
			+ \left( \beta^2 - \frac{1}{16 \beta^2} \right)
				\sinh(s - \I \theta),
	\\
	\label{eq:summary-punc-t}
	Z_4^{(t)}(s, \theta)
		&
		= - \frac{8\beta^2
				+ (1 + 16 \beta^4) \cosh(s - \I \theta)
				- (1 - 16 \beta^4) \sinh(s - \I \theta)}
				{8\beta^2
				- (1 + 16 \beta^4) \cosh(s - \I \theta)
				+ (1 - 16 \beta^4) \sinh(s - \I \theta)},
	\\
	\label{eq:summary-punc-u}
	Z_4^{(u)}(s, \theta)
		&
		= - \frac{16 \beta^2}
				{8\beta^2
				- (1 + 16 \beta^4) \cosh(s - \I \theta)
				+ (1 - 16 \beta^4) \sinh(s - \I \theta)}.
\end{align}
\end{subequations}

\begin{figure}[ht]
	\centering
	\includegraphics[scale=1.5]{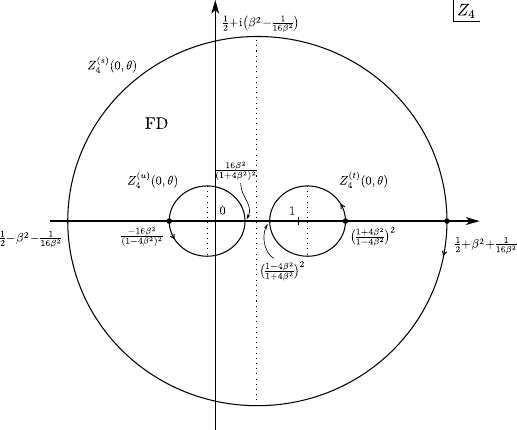}
	\caption{%
		Boundary of the $s$-, $t$- and $u$-channels.
		The dot on the curves locate the point $\theta = 0$ and the arrow denotes the direction of increasing $\theta$.
	}
	\label{fig:M04-channels}
\end{figure}

The boundaries of the different channels, $\pd \mc F_{0,4}^{(s,t,u)}$, correspond to $\abs{q} = 1$ or $s = 0$ (\Cref{fig:M04-channels}):
\begin{subequations}
\label{eq:summary-bdy}
\begin{align}
	\label{eq:summary-bdy-s}
	Z_4^{(s)}(0, \theta)
		= \frac{1}{2}
				+ \left( \beta^2 + \frac{1}{16 \beta^2} \right)
					\cos \theta
				- \I \left( \beta^2 - \frac{1}{16 \beta^2} \right)
					\sin \theta,
	\\
	\label{eq:summary-bdy-t}
	Z_4^{(t)}(0, \theta)
		= - \frac{8\beta^2
				+ (1 + 16 \beta^4) \cos \theta
				+ \I (1 - 16 \beta^4) \sin \theta}
				{8\beta^2
				- (1 + 16 \beta^4) \cos \theta
				- \I (1 - 16 \beta^4) \sin \theta},
	\\
	\label{eq:summary-bdy-u}
	Z_4^{(u)}(0, \theta)
		= - \frac{16 \beta^2}
				{8\beta^2
				- (1 + 16 \beta^4) \cos \theta
				- \I (1 - 16 \beta^4) \sin \theta}.
\end{align}
\end{subequations}
We have:
\begin{equation}
	\begin{gathered}
	\abs{Z_4^{(s)}(0, \theta)}
		= \beta^2 + \frac{1}{16 \beta^2} + \frac{\cos \theta}{2},
	\\
	\abs{Z_4^{(u)}(0, \theta)}
		= \frac{1}{\abs{Z_4^{(s)}(0, \theta + \pi)}},
	\qquad
	\abs{Z_4^{(u)}(0, \theta)}
		= \frac{\abs{Z_4^{(s)}(0, \theta)}}{\abs{Z_4^{(s)}(0, \theta + \pi)}}.
	\end{gathered}
\end{equation}
In the $s$-channel, we can also write:
\begin{equation}
	\label{eq:bdy-s-exp}
	Z_4^{(s)}(0, \theta)
		= \frac{1}{2}
			+ \beta^2 \, \e^{- \I \theta}
			+ \frac{\e^{\I \theta}}{16 \beta^2}.
\end{equation}

We must ensure that coordinate patches do not overlap: this is the case if
\begin{equation}
	\label{eq:S04:bound-beta}
	\beta
		\ge \beta_{0,4}
		= \frac{1}{2} + \frac{1}{\sqrt{2}}
		\approx \num{1.207}.
\end{equation}
which is automatically satisfied since $\beta_{0,4} < \beta_{0,3}$ given in \eqref{eq:S03:bound-beta}.
In the large stub limit $\beta \gg 1$, the boundaries all approach circles:
\begin{equation}
	Z_4^{(s)}(0, \theta)
		\sim \frac{1}{2} + \beta^2 \, \e^{- \I\theta},
	\qquad
	Z_4^{(t)}(0, \theta)
		\sim 1 + \frac{\e^{\I\theta}}{\beta^2},
	\qquad
	Z_4^{(u)}(0, \theta)
		\sim - \frac{\e^{\I\theta}}{\beta^2}.
\end{equation}

The formulas \eqref{eq:summary-bdy} are well-adapted to determine the geometrical shapes of the boundaries.
From the formulas in \Cref{sec:ellipses}, we see that the boundary of the $s$-channel is an ellipse with parameters
\begin{equation}
	\begin{gathered}
	z_c
		= \frac{1}{2},
	\qquad
	\theta_0
		= 0,
	\qquad
	a
		= \beta^2 + \frac{1}{16 \beta^2},
	\qquad
	b
		= \beta^2 - \frac{1}{16 \beta^2},
	\\
	e
		= \frac{8 \beta^2}{1 + 16 \beta^4},
	\qquad
	c
		= \frac{1}{2},
	\qquad
	F_1
		= 0,
	\qquad
	F_2
		= 1.
	\end{gathered}
\end{equation}
Moreover, the direction of $\theta$ is reversed with respect to the canonical orientation.
For $\beta = \beta_{03} = 3/2$, the eccentricity is
\begin{equation}
	e(\beta = 3/2)
		 \approx 0.22,
\end{equation}
and $e$ decreases rapidly as $\beta$ increases.
Note the following relation:
\begin{equation}
	\sqrt{\frac{1 + e}{1 - e}}
		= \frac{1 + 4 \beta^2}{1 - 4 \beta^2}.
\end{equation}

The $t$- and $u$-channels correspond to inversions of the $s$-channel ellipse with respect to each focus: this implies that the curves are convex limaçons of Pascal with parameters (see \Cref{sec:ellipses})
\begin{equation}
	\beta_\ell
		= \frac{1}{a (1 - e^2)},
	\qquad
	\alpha_\ell
		= \frac{e}{a (1 - e^2)}.
\end{equation}
The distance of a point to the origin in terms of the angle $\nu$ is:
\begin{equation}
	\rho(\nu)
		= \frac{1}{a (1 - e^2)} \left( - \frac{e}{2} + \e^{\I \nu} - \frac{\e^{2 \I \nu}}{2} \right).
\end{equation}

Finally, we can determine the volumes of the $t$- and $u$-channel regions:
\begin{equation}
	\mathrm{Vol}\big(\mc F_{0,4}^{(t)}\big)
		= \mathrm{Vol}\big(\mc F_{0,4}^{(u)}\big)
		= 256 \pi \beta^4 \,
			\frac{1 + 64 \beta^2 + 256 \beta^8}{(1 - 16 \beta^4)^4}
\end{equation}
and of the complement of the $s$-channel region:
\begin{equation}
	\mathrm{Vol}\big(\mc M_{0,4} - \mc F_{0,4}^{(s)}\big)
		= \pi \left( \beta^4 - \frac{1}{256 \beta^4} \right)
\end{equation}
using formulas from \eqref{sec:ellipses}.
We thus get the volume of the fundamental vertex region:
\begin{equation}
	\begin{aligned}
	\mathrm{Vol}\big(\mc V_{0,4}\big)
		&
		= \mathrm{Vol}\big(\mc M_{0,4} - \mc F_{0,4}^{(s)}\big)
			- \mathrm{Vol}\big(\mc F_{0,4}^{(t)}\big)
			- \mathrm{Vol}\big(\mc F_{0,4}^{(u)}\big)
		\\ &
		= \pi \beta^4
			- \frac{513 \pi}{256 \beta^4} + O(\beta^{-8}).
	\end{aligned}
\end{equation}
For $\beta = 1.5$, we have $\mathrm{Vol}\big(\mc V_{0,4}\big) = \num{14.5239}$.

\section{Quartic vertex}
\label{sec:vertex}

The $4$-point vertex $\mc V_{0,4}$ is defined first by the part of the moduli space not covered by the factorized graphs (white part \Cref{fig:simple:M04-prop}).
We need to find a parametrization of the vertex region, but also local coordinates $F^{(\times)}_a(W, Z_4)$ which match the ones of the factorized graphs on the common boundaries (in the rest of this section, we omit the superscript on the maps for the vertex region).
To simplify the problem, we will work inside the fundamental domain for $S_4$ described in \Cref{sec:S04:fundamental-domain}.

\subsection{Vertex region parametrization}

\begin{figure}[ht]
	\centering
	\includegraphics[scale=1.5]{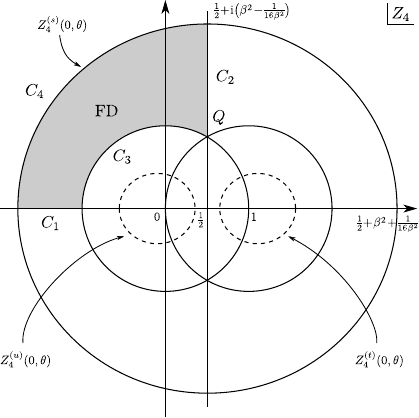}
	\caption{%
		Fundamental domain of $S_4$ (FD, in grey) inside the fundamental vertex region $\mc V_{0,4}$.
		The images of the unit circle under the $S_3$ maps are also displayed.
	}
	\label{fig:FD-V04}
\end{figure}

The $S_4$ fundamental domain of the vertex region is obtained by excluding the $s$-channel region from the fundamental domain \eqref{eq:fd-cond-C}:
\begin{equation}
	\label{eq:fd-cond-V04}
	\Re Z_4
		\le \frac{1}{2},
	\qquad
	\abs{Z_4}
		\ge 1,
	\qquad
	\Im Z_4
		\ge 0,
	\qquad
	\abs{Z_4}
		\le \abs{Z_4^{(s)}(0, \theta)}
\end{equation}
for all $\theta \in [0, 2\pi)$, see \eqref{eq:summary-bdy-s}.
Hence, it is delimited by the following curves (\Cref{fig:FD-V04}):
\begin{equation}
	\label{eq:fd-cond-V04-bdy}
	\begin{gathered}
	C_1:
		\Re Z_4
			= \frac{1}{2},
	\qquad
	C_2:
		\Im Z_4
			= 0,
	\\
	C_3:
		\abs{Z_4}
			= 1,
	\qquad
	C_4:
		\abs{Z_4}
			= \abs{Z_4^{(s)}(0, \theta)}
	\end{gathered}
\end{equation}

\begin{figure}[ht]
	\centering
	\includegraphics[scale=1.5]{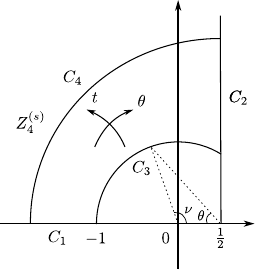}
	\caption{%
		Interpolating coordinates $(t, \theta)$ for the fundamental domain of the vertex region.
	}
	\label{fig:FD-S4-interp}
\end{figure}

We can parametrize the vertex region by interpolating radially between the boundary $C_3$ and $C_4$ with a parameter $t \in [0, 1]$, and describing the polar direction with the angle $\theta$ of the $s$-channel ellipse (\Cref{fig:FD-S4-interp}).
The unit circle centered at $Z_4 = 0$ can be parametrized as
\begin{equation}
	Z_4^{(\circ)}(\theta)
		:= \frac{1}{2} - \rho(\theta) \, \e^{- \I \theta},
	\qquad
	\rho(\theta)
		= \frac{1}{2} \Bigl( \cos \theta + \sqrt{\cos^2 \theta + 3} \Bigr).
\end{equation}
It is easy to check that $Z_4^{(\circ)}(0) = - 1$ and $Z_4^{(\circ)}(\pi / 2) = Q$.
Then, we just need to do a linear interpolation:
\begin{equation}
	\label{eq:vertex-Z4}
	\boxed{
	Z_4(t, \theta)
		:= (1 - t) Z_4^{(\circ)}(\theta)
			+ t \, Z_4^{(s)}(0, \theta + \pi),
	\qquad
	t \in [0, 1].
	}
\end{equation}

We also note that the following Joukowski transformation maps the unit circle to the $s$-channel ellipse:
\begin{equation}
	J(z)
		= \frac{1}{2}
			+ \beta^2 \, z
			+ \frac{1}{16 \beta^2 \, z},
\end{equation}
see \eqref{eq:bdy-s-exp}.
This induces a map in the $z$-plane, which is conformal but not $\group{SL}(2, \C)$.
Hence, it cannot be used to determine local coordinates on $C_3$ by applying its inverse on local coordinates on the $s$-channel boundary (also note that the map converts the angle $\nu$ to the angle $\theta$, so angles are misaligned for the fundamental domain).

\subsection{Constraints on boundary}
\label{sec:vertex:constraints}

We first impose that the local coordinates for moduli $Z_4$ and $\bar Z_4$ are related by complex conjugation:
\begin{equation}
	F_i(W, \bar Z_4)
		= \overline{F_i(\bar W, Z_4)}.
\end{equation}

Next, we have to impose equality between the local coordinates obtained by different elements of $S_4$ but with identical images of $Z_4$.
Taking into account that local coordinates are permuted differently, \eqref{eq:map-coord-S4-gen} and \eqref{eq:map-coord-S4-S3} give the relations:
\begin{equation}
	\label{eq:Fi-constraints-perm}
	\begin{gathered}
	S_{12} \circ F_1(W, Z_4)
		= S_{34} \circ F_2(W, Z_4),
	\qquad
	S_{13} \circ F_1(W, Z_4)
		= S_{24} \circ F_3(W, Z_4),
	\\
	S_{23} \circ F_1(W, Z_4)
		= S_{14} \circ F_4(W, Z_4).
	\end{gathered}
\end{equation}
This means that one can obtain $F_2$, $F_3$ and $F_4$ from $F_1$.
This is similar to what happened in \eqref{sec:setup:S03}, where the invariance under $S_3$ was sufficient to give $f_1$ and $f_\infty$ from $f_0$.

\medskip

We can obtain constraints on $F_1$ by looking at the different boundaries of the FD: local coordinates on each side of each boundary must coincide on the latter.
If the permutation map sends the modulus in the lower half-plane, then the reality condition can be used to rewrite it in the upper half-plane.
We find:
\begin{enumerate}
	\item On $C_1$, $\Im Z_4 = 0$:
	\begin{equation}
		F_i(W, Z_4)
			= \overline{F_i(\bar W, \bar Z_4)}
	\end{equation}
	for $Z_4 = x \in \R$.

	\item On $C_2$,  $\Re Z_4 = 1/2$:
	\begin{equation}
		F_1(W, \bar Z_4)
			= S_{24} \circ F_1(W, Z_4)
			= \frac{F_1(W, Z_4)}{Z_4}
	\end{equation}
	for $Z_4 = \frac{1}{2} + \I y, y \in \R$.

	\item On $C_3$, $\abs{Z_4} = 1$:
	\begin{equation}
		F_1(W, \bar Z_4)
			= S_{34} \circ F_1(W, Z_4)
			= \frac{F_1(W, Z_4) \, (1 - Z_4)}{F_1(W, Z_4) - Z_4}
	\end{equation}
	for $Z_4 = \e^{\I \nu}, \nu \in [0, 2\pi)$.

	\item On $C_4$, the boundary with the $s$-channel $Z_4 = Z_4^{(s)}(q)$:
	\begin{equation}
		F_1\Big( W, Z_4^{(s)}(q) \Big)
			= F_1^{(s)}\Big( W, Z_4^{(s)}(q) \Big)
	\end{equation}
	for $q = \e^{\I \theta}, \theta \in [0, 2\pi)$.
\end{enumerate}
The motivation for using the maps $S_{24}$ (induces $g_{0,1}(Z_4)$ which lands on the right of $C_2$ after complex conjugation) and $S_{34}$ (induces $g_{0,\infty}(Z_4)$ which lands below $C_3$ after complex conjugation) is that they leave the first puncture invariant, such that we can match $F_1$ with its transformation without any permutation.

When solving for the constraints, we have to allow for an $Z_4$-dependent rescaling of the local coordinate because we did not normalize the $\group{SL}(2, \C)$ coefficients:
\begin{equation}
	F_1'(W, Z_4')
		= F_1(\lambda  \, W, Z_4).
\end{equation}

\medskip

We parametrize $F_1$ as
\begin{equation}
	F_1(W, Z_4)
		= \frac{A(Z_4, \bar Z_4) W}{C(Z_4, \bar Z_4) W + D(Z_4, \bar Z_4)}.
\end{equation}
There is no constant term in the numerator since $F_1(0) = 0$.
In general, we will omit the dependence in $\bar Z_4$.

The strategy to build the local coordinates is as follows:
\begin{enumerate}
	\item Find local coordinates on $C_3$ by solving the constraints, denoting the coefficients as $A^{(\circ)}$, etc.

	\item Check that the previous coordinates solve the constraints on the corners $C_1 \cup C_3$ and $C_2 \cup C_3$.

	\item Write the coefficients of the functions as an interpolation between the coefficients on $C_3$ and $C_4$:
	\begin{equation}
		A(t, \theta)
			= (1 - t) \, A^{(\circ)}(\theta) + t \, A^{(s)}(0, \theta),
		\qquad
		\text{etc.}
	\end{equation}

	\item Check that one recovers the modulus $Z_4(t, \theta)$ in \eqref{eq:vertex-Z4} when evaluating $F_4(0)$.
\end{enumerate}
In this paper, we do not provide an explicit solution to the constraints (since it is not expected to be unique).

\paragraph{Constraint on $C_1$}

The equation for $Z_4 = x \in \R$ is:
\begin{equation}
	\frac{\lambda_1 \, \overline{A(x)} W}
			{\lambda_1 \, \overline{C(x)} W + \overline{D(x)}}
		= \frac{A(x) W}{C(x) W + D(x)}
\end{equation}
such that
\begin{equation}
	\label{eq:V04-constraint-C1}
	\lambda_1 \, \overline{A(x)}
		= A(x),
	\qquad
	\lambda_1 \, \overline{C(x)}
		= C(x),
	\qquad
	\overline{D(x)}
		= D(x).
\end{equation}

\paragraph{Constraint on $C_2$}

The equation for $Z_4 = 1/2 + \I y \in \R$ is:
\begin{equation}
	\frac{\lambda_2 \, (1/2 + \I y) A(1/2 - \I y) W}{\lambda_2 \, C(1/2 - \I y) W + D(1/2 - \I y)}
		= \frac{A(1/2 + \I y) W}{C(1/2 + \I y) W + D(1/2 + \I y)}
\end{equation}
such that
\begin{equation}
	\label{eq:V04-constraint-C2}
	\begin{aligned}
		\lambda_2 \, (1/2 + \I y) A(1/2 - \I y)
		&
		= A(1/2 + \I y),
	\\
	\lambda_2 \, C(1/2 - \I y)
		&
		= C(1/2 + \I y),
	\\
	D(1/2 - \I y)
		&
		= D(1/2 + \I y).
	\end{aligned}
\end{equation}

\paragraph{Constraint on $C_3$}

The equation for $Z_4 = \e^{\I \nu}, \nu \in [0, 2\pi)$ is:
\begin{equation}
	\frac{\lambda_3 \, A(\e^{- \I \nu}) W}{\lambda_3 \, C(\e^{- \I \nu}) W + D(\e^{- \I \nu})}
		= \frac{(1 - \e^{\I \nu}) A(\e^{\I \nu}) W}{[A(\e^{\I \nu}) - \e^{\I \nu} C(\e^{\I \nu})] W - \e^{\I \nu} D(\e^{\I \nu})}
\end{equation}
such that
\begin{equation}
	\label{eq:V04-constraint-C3}
	\begin{aligned}
	\lambda_3 \, A(\e^{- \I \nu})
		&
		= (1 - \e^{\I \nu}) A(\e^{\I \nu}),
	\\
	\lambda_3 \, C(\e^{- \I \nu})
		&
		= A(\e^{\I \nu}) - \e^{\I \nu} C(\e^{\I \nu}),
	\\
	D(\e^{- \I \nu})
		&
		= - \e^{\I \nu} D(\e^{\I \nu}).
	\end{aligned}
\end{equation}
It seems simpler to solve this equation with the angle $\nu$ for the center $Z_4 = 0$, but later we will need to convert it to the angle $\theta$ measuring around $Z_4 = 1/2$.

\paragraph{Constraint on $C_4$}

The equation at the boundary with the $s$-channel $Z_4 = Z_4^{(s)}(\e^{\I \theta}), \theta \in [0, 2\pi)$ gives
\begin{equation}
	\frac{\lambda_4 \, A\big(Z_4^{(s)}(\e^{\I\theta})\big) W}{\lambda_4 \, C\big(Z_4^{(s)}(\e^{\I\theta})\big) W + D\big(Z_4^{(s)}(\e^{\I\theta})\big)}
		= \frac{2 (4 \beta^2 + \e^{\I\theta}) W}{(3 \e^{\I\theta} + 4 \beta^2) W + 2 \beta (4 \beta^2 - \e^{\I\theta})}.
\end{equation}
such that
\begin{equation}
	\label{eq:V04-constraint-C4}
	\begin{aligned}
	\lambda_4 \, A\big(Z_4^{(s)}(\e^{\I\theta})\big)
		&
		= 2 (4 \beta^2 + \e^{\I\theta}),
	\\
	\lambda_4 \, C\big(Z_4^{(s)}(\e^{\I\theta})\big)
		&
		= 3 \e^{\I\theta} + 4 \beta^2,
	\\
	D\big(Z_4^{(s)}(\e^{\I\theta})\big)
		&
		= 2 \beta (4 \beta^2 - \e^{\I\theta}).
	\end{aligned}
\end{equation}

\paragraph{Constraints on points}

We need to specify the constraints at the corner points $Z_4 = -1$ ($C_1 \cap C_3$) and $Z_4 = Q$ ($C_2 \cap C_3$).

For $Z_4 = - 1$ ($\nu = \pi$, $x = -1$), we find from \eqref{eq:V04-constraint-C1} and \eqref{eq:V04-constraint-C3}:
\begin{equation}
	\label{eq:V04-constraint-C1C3}
	\begin{gathered}
	\lambda_1 \, \overline{A(-1)}
		= A(-1),
	\qquad
	\lambda_1 \, \overline{C(-1)}
		= C(-1),
	\\
	\overline{D(-1)}
		= D(-1),
	\qquad
	\lambda_3 \, A(-1)
		= 2 A(-1),
	\\
	\lambda_3 \, C(-1)
		= A(-1) + C(-1),
	\qquad
	D(-1)
		= D(-1).
	\end{gathered}
\end{equation}
This implies $\lambda_3 = 2$ (for $Z_4 = -1$), $C(-1) = A(-1)$.

For $Z_4 = Q$ ($\nu = \pi/3$ and $y = \sqrt{3}/2$), we find from \eqref{eq:V04-constraint-C2} and \eqref{eq:V04-constraint-C3}:
\begin{equation}
	\label{eq:V04-constraint-C2C3}
	\begin{gathered}
	\lambda_2 \, Q \, A(\bar Q)
		= A(Q),
	\qquad
	\lambda_2 \, C(\bar Q)
		= C(Q),
	\\
	\lambda_2 \, D(\bar Q)
		= D(Q),
	\qquad
	\lambda_3 \, A(\bar Q)
		= \bar Q A(Q),
	\\
	\lambda_3 \, C(\bar Q)
		= A(Q) - Q C(Q),
	\qquad
	D(\bar Q)
		= - Q D(Q).
	\end{gathered}
\end{equation}
We have used that $\bar Q = 1/Q = 1 - Q$.

\section{Discussions}

In this paper, we have described the decomposition of the moduli space of 4-punctured spheres into Feynman and vertex regions induced by the cubic $\group{SL}(2, \C)$ vertex.
The formulas for the boundaries of the different regions and of their volumes are analytical, which is a major simplification over minimal area vertices.
We have determined a fundamental region under the permutation of punctures for the vertex region, and determined the constraints which must be satisfied by the local coordinates on the boundaries of this domain.
In particular, we have obtained an explicit parametrization of the fundamental domain.
However, we did not succeed in finding an explicit example of local coordinates.

In future works, we plan to come back on this issue, and extend the computations to the 2-punctured torus, which would allow computing mass renormalization~\cite{Pius:2014:MassRenormalizationStringGeneral,Pius:2014:MassRenormalizationStringSpecial}.
It would also be interesting to see if there is any way to determine general properties of higher-order vertices: indeed, since the maps are all $\group{SL}(2, \C)$, the maps obtained after gluing should all display properties similar to those of this paper.

\section*{Acknowledgments}

We would like to thank Atakan Hilmi Fırat for reading the draft, Ted Erler for his support at early stages of this project; Ivo Sachs, Ashoke Sen and Jakub Vošmera for discussions.

This project has received funding from the European Union's Horizon 2020 research and innovation program under the Marie Skłodowska-Curie grant agreement No 891169.
This work is supported by the National Science Foundation under Cooperative Agreement PHY-2019786 (The NSF AI Institute for Artificial Intelligence and Fundamental Interactions, \url{http://iaifi.org/}).
SM would like to thank CEICO (Prague) for its hospitality while this work was carried out.

\appendix

\section{\texorpdfstring{$\mathrm{SL}(2, \C)$}{SL(2, C)} maps}
\label{app:sl2c}

A $\group{SL}(2, \C)$ transformation is a function
\begin{equation}
	g(z)
		= \frac{a z + b}{c z + d},
	\qquad
	a d - b c
		= 1,
\end{equation}
where $a, b, c, d \in \C$.
Since $\dim_\C \group{SL}(2, C) = 3$, one needs three conditions to completely fix $g(z)$: this also means that any three points $z_1$, $z_2$ and $z_3$ can be set to a convenient value with such a transformation.
Given three points $z_1$, $z_2$ and $z_3$, the $\group{SL}(2, \C)$ function which generates a permutation $z_{\sigma(i)} = \sigma(z_i)$ is denoted by
\begin{equation}
	g_{\sigma}(z)
		= g_{\sigma(1)\sigma(2)\sigma(3)}(z).
\end{equation}

If the permutation leaves one puncture $z_1$ fixed and exchanges the two others $z_2$ and $z_3$ -- that is, $\sigma(1) = 1$, $\sigma(2) = 3$, $(\sigma(3) = 2$ --, we abbreviate the function as
\begin{equation}
	g_{23}(z)
		:= g_{132}(z).
\end{equation}
By solving the equations
\begin{equation}
	g_{23}(z_1)
		= z_1,
	\qquad
	g_{23}(z_2)
		= z_3,
	\qquad
	g_{23}(z_3)
		= z_2,
\end{equation}
we find
\begin{equation}
	g_{23}(z)
		= \frac{z (z_2 z_3 - z_1^2) + z_1 \big( z_1 (z_2 + z_3) - 2 z_2 z_3 \big)}{z (2 z_1 - z_2 - z_3) + z_2 z_3}.
\end{equation}

The group $S_3 \subset \group{SL}(2, \C)$ of the permutations of the three punctures $(0, 1, \infty)$ is generated by
\begin{equation}
	\label{eq:sl2c:S3-perm}
	g_{01}(z)
		= 1 - z,
	\qquad
	g_{0\infty}(z)
		= \frac{1}{z}.
\end{equation}
Other permutations can be generated by composition:
\begin{equation}
	g_{1\infty}(z)
		= g_{01} \circ g_{0\infty} \circ g_{01}(z)
		= \frac{z}{z - 1},
	\qquad
	g_{1\infty 0}(z)
		= g_{0\infty} \circ g_{01}(z)
		= \frac{1}{1 - z}.
\end{equation}

\section{Ellipses and related curves}
\label{sec:ellipses}

\subsection{Ellipses}

\begin{figure}[ht]
	\centering
	\includegraphics[scale=2]{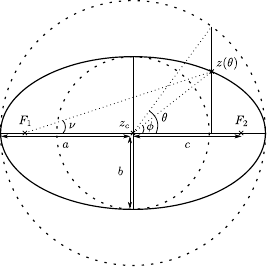}
	\caption{Ellipse with semi-axis $a$ and $b$ centered at $z_c$.}
	\label{fig:ellipse}
\end{figure}

An ellipse is parametrized by (\Cref{fig:ellipse}).
\begin{equation}
	z(\theta)
		= z_c + a \cos(\theta + \theta_0) + \I b \sin(\theta + \theta_0),
\end{equation}
where $z_c$ is the center, $a$ and $b$ are the major and minor semi-axis, $\theta_0$ a phase shift (origin of the angle).
The angle $\theta$ (“eccentric anomaly”) measures the angle between the horizontal axis and the vertical projection of the point on the circle of radius $a$ centered at $z_c$.
The eccentricity is defined as
\begin{equation}
	e
		= \sqrt{1 - \left( \frac{b}{a} \right)^2},
\end{equation}
from which the distance between each focus to the center can be read:
\begin{equation}
	c
		= \sqrt{a^2 - b^2}
		= a e.
\end{equation}

The area of the ellipse is
\begin{equation}
	A
		= \pi a b.
\end{equation}

It is possible to introduce two additional angles $\nu$ (“true anomaly”) and $\phi$ between the lines joining the focus $F_1$ and the center to $z(\theta)$ and the real axis:
\begin{equation}
	\label{eq:map-angles-ellipse}
	\tan \frac{\nu}{2}
		= \sqrt{\frac{1 + e}{1 - e}} \, \tan \frac{\theta}{2},
	\qquad
	\cos \phi
		= \frac{a \cos \theta}{\sqrt{a^2 \cos^2 \theta + b^2 \sin^2 \theta}}.
\end{equation}
These new angles are useful because they are more uniform on the ellipse.
The distances $\rho(\nu)$ and $r(\phi)$ from the point $z(\theta)$ to the focus $F_1$ and to the center are
\begin{subequations}
\label{eq:ellipse-distances}
\begin{align}
	\rho(\nu)
		&
		= \frac{a (1 - e^2)}{1 - e \cos \nu},
	\\
	r(\phi)
		&
		= \frac{a b}{\sqrt{b^2 \cos^2 \phi + a^2 \sin^2 \phi}}
		= \frac{b}{\sqrt{1 - e^2 \cos^2 \phi}}.
\end{align}
\end{subequations}

\subsection{Pascal limaçon}

The inverse of an ellipse with respect to one of its focus gives a limaçon of Pascal (general Möbius transformations of ellipses are studied in~\cite{Coffman:2007:MobiusTransformationsEllipses}.).
We focus on the case of a focus centered at the origin.
The distance of a point of a limaçon with parameters $\alpha$ and $\beta$ to the origin reads:
\begin{equation}
	\rho(\nu)^{-1}
		= \beta - \alpha \cos \nu,
\end{equation}
where $\nu$ is the angle between the point and the real axis.
The canonical parametrization of a limaçon is found by multiplying this with $\e^{\I\nu}$:
\begin{equation}
	z(\nu)
		= (\beta - \alpha \cos \nu) \e^{\I \nu}
		= - \frac{\alpha}{2} + \beta \e^{\I \nu} - \frac{\alpha}{2} \, \e^{2 \I \nu}.
\end{equation}
To find the relation between the parameters of an ellipse with one focus at the origin and the limaçon obtained by taking the inverse with respect to this focus, one needs to use the expression $\rho(\nu)$ in \eqref{eq:ellipse-distances}.
We find:
\begin{equation}
	\label{eq:limacon-param-ellipse}
	\beta
		= \frac{1}{a (1 - e^2)},
	\qquad
	\alpha
		= \frac{e}{a (1 - e^2)}.
\end{equation}
For $\alpha = \beta$, the curve becomes a cardioid.
The limaçon is convex if
\begin{equation}
	\beta
		> 2 \alpha
	\quad \Longrightarrow \quad
	e
		< \frac{1}{2}.
\end{equation}
This will always hold in our case.
Finally, the area of the limaçon is
\begin{equation}
	A
		= \pi \left( \frac{\alpha^2}{2} + \beta^2 \right).
\end{equation}
For the limaçon defined by \eqref{eq:limacon-param-ellipse}, we find:
\begin{equation}
	A
		= \frac{\pi}{2} \, \frac{2 + e^2}{a^2 (1 - e^2)^2}
		= \frac{\pi}{2} \, \frac{3 a^2 - b^2}{b^4}.
\end{equation}

\printbibliography[heading=bibintoc]

@article{Chiaffrino:2021:QFTStubs,
  title = {{{QFT}} with {{Stubs}}},
  author = {Chiaffrino, Christoph and Sachs, Ivo},
  year = {2021},
  month = aug,
  journal = {arXiv:2108.04312 [hep-th]},
  eprint = {2108.04312},
  primaryclass = {hep-th},
  url = {http://arxiv.org/abs/2108.04312},
  urldate = {2021-08-11},
  archiveprefix = {arxiv}
}

@article{Coffman:2007:MobiusTransformationsEllipses,
  title = {Möbius Transformations and Ellipses},
  author = {Coffman, Adam and Frantz, Marc},
  year = {2007},
  journal = {Pi Mu Epsilon Journal},
  volume = {12},
  number = {6},
  eprint = {24340762},
  eprinttype = {jstor},
  pages = {339--345},
  issn = {0031-952X},
  url = {https://www.jstor.org/stable/24340762},
  urldate = {2019-09-03}
}

@article{Costello:2019:HyperbolicStringVertices,
  title = {Hyperbolic {{String Vertices}}},
  author = {Costello, Kevin and Zwiebach, Barton},
  year = {2019},
  month = aug,
  eprint = {1909.00033},
  url = {https://arxiv.org/abs/1909.00033},
  urldate = {2019-09-05},
  archiveprefix = {arxiv},
  langid = {english}
}

@misc{Eniceicu:2022:ZZAnnulusOnepoint,
  title = {The {{ZZ}} Annulus One-Point Function in Non-Critical String Theory: {{A}} String Field Theory Analysis},
  shorttitle = {The {{ZZ}} Annulus One-Point Function in Non-Critical String Theory},
  author = {Eniceicu, Dan Stefan and Mahajan, Raghu and Maity, Pronobesh and Murdia, Chitraang and Sen, Ashoke},
  year = {2022},
  month = oct,
  number = {arXiv:2210.11473},
  eprint = {2210.11473},
  primaryclass = {hep-th},
  publisher = {{arXiv}},
  doi = {10.48550/arXiv.2210.11473},
  urldate = {2022-10-23},
  archiveprefix = {arxiv}
}

@misc{Erbin:2018:CubicClosedString,
  title = {Towards a Cubic Closed String Field Theory?},
  author = {Erbin, Harold},
  year = {2018},
  month = jul,
  address = {{Ringberg Castle}},
  url = {https://www.theorie.physik.uni-muenchen.de/activities/workshops/archive_workshops_conferences/ringberg_geometry_strings_18/slides_ringberg_2018/index.html}
}

@book{Erbin:2021:StringFieldTheory,
  title = {String {{Field Theory}}: {{A Modern Introduction}}},
  shorttitle = {String {{Field Theory}}},
  author = {Erbin, Harold},
  year = {2021},
  month = mar,
  series = {Lecture {{Notes}} in {{Physics}}},
  eprint = {2301.01686},
  publisher = {{Springer}},
  doi = {10.1007/978-3-030-65321-7},
  urldate = {2021-04-04},
  archiveprefix = {arxiv},
  isbn = {978-3-030-65320-0},
  langid = {english}
}

@misc{Erbin:2022:Characterizing4stringContact,
  title = {Characterizing 4-String Contact Interaction Using Machine Learning},
  author = {Erbin, Harold and Fırat, Atakan Hilmi},
  year = {2022},
  month = nov,
  number = {arXiv:2211.09129},
  eprint = {2211.09129},
  primaryclass = {hep-th},
  publisher = {{arXiv}},
  doi = {10.48550/arXiv.2211.09129},
  urldate = {2022-11-21},
  archiveprefix = {arxiv}
}

@misc{Erbin:2023:OpenStringStub,
  title = {Open String Stub as an Auxiliary String Field},
  author = {Erbin, Harold and Fırat, Atakan Hilmi},
  year = {2023},
  month = aug,
  number = {arXiv:2308.08587},
  eprint = {2308.08587},
  primaryclass = {hep-th},
  publisher = {{arXiv}},
  doi = {10.48550/arXiv.2308.08587},
  urldate = {2023-08-21},
  archiveprefix = {arxiv}
}

@article{Erler:2017:OneLoopTadpole,
  title = {One {{Loop Tadpole}} in {{Heterotic String Field Theory}}},
  author = {Erler, Theodore and Konopka, Sebastian and Sachs, Ivo},
  year = {2017},
  journal = {Journal of High Energy Physics},
  volume = {11},
  eprint = {1704.01210},
  pages = {056},
  doi = {10.1007/JHEP11(2017)056},
  urldate = {2017-11-22},
  archiveprefix = {arxiv}
}

@article{Erler:2020:FourLecturesClosed,
  title = {Four {{Lectures}} on {{Closed String Field Theory}}},
  author = {Erler, Theodore},
  year = {2020},
  month = jan,
  journal = {Physics Reports},
  eprint = {1905.06785},
  pages = {S0370157320300132},
  issn = {03701573},
  doi = {10.1016/j.physrep.2020.01.003},
  urldate = {2020-02-05},
  archiveprefix = {arxiv}
}

@misc{Erler:2022:FourLecturesAnalytic,
  title = {Four {{Lectures}} on {{Analytic Solutions}} in {{Open String Field Theory}}},
  author = {Erler, Theodore},
  year = {2022},
  month = jul,
  number = {arXiv:1912.00521},
  eprint = {1912.00521},
  primaryclass = {hep-th},
  publisher = {{arXiv}},
  doi = {10.48550/arXiv.1912.00521},
  urldate = {2022-07-28},
  archiveprefix = {arxiv}
}

@article{Firat:2021:HyperbolicThreestringVertex,
  title = {Hyperbolic Three-String Vertex},
  author = {Fırat, Atakan Hilmi},
  year = {2021},
  month = aug,
  journal = {Journal of High Energy Physics},
  volume = {2021},
  number = {8},
  eprint = {2102.03936},
  primaryclass = {hep-th},
  pages = {35},
  issn = {1029-8479},
  doi = {10.1007/JHEP08(2021)035},
  urldate = {2022-09-09},
  archiveprefix = {arxiv}
}

@misc{Firat:2023:BootstrappingClosedString,
  title = {Bootstrapping Closed String Field Theory},
  author = {Fırat, Atakan Hilmi},
  year = {2023},
  month = feb,
  number = {arXiv:2302.12843},
  eprint = {2302.12843},
  primaryclass = {hep-th},
  publisher = {{arXiv}},
  doi = {10.48550/arXiv.2302.12843},
  urldate = {2023-02-28},
  archiveprefix = {arxiv}
}

@misc{Firat:2023:HyperbolicStringTadpole,
  title = {Hyperbolic String Tadpole},
  author = {Fırat, Atakan Hilmi},
  year = {2023},
  month = jun,
  number = {arXiv:2306.08599},
  eprint = {2306.08599},
  primaryclass = {hep-th},
  publisher = {{arXiv}},
  url = {http://arxiv.org/abs/2306.08599},
  urldate = {2023-06-16},
  archiveprefix = {arxiv}
}

@article{Hubbard:1959:CalculationPartitionFunctions,
  title = {Calculation of {{Partition Functions}}},
  author = {Hubbard, J.},
  year = {1959},
  month = jul,
  journal = {Physical Review Letters},
  volume = {3},
  number = {2},
  pages = {77--78},
  doi = {10.1103/PhysRevLett.3.77},
  urldate = {2018-07-12}
}

@misc{Maccaferri:2022:ClassicalCosmologicalConstant,
  title = {The Classical Cosmological Constant of Open-Closed String Field Theory},
  author = {Maccaferri, Carlo and Vošmera, Jakub},
  year = {2022},
  month = jul,
  number = {arXiv:2208.00410},
  eprint = {2208.00410},
  primaryclass = {hep-th},
  publisher = {{arXiv}},
  doi = {10.48550/arXiv.2208.00410},
  urldate = {2022-08-03},
  archiveprefix = {arxiv}
}

@article{Moeller:2004:ClosedBosonicString,
  title = {Closed {{Bosonic String Field Theory At Quartic Order}}},
  author = {Moeller, Nicolas},
  year = {2004},
  month = nov,
  journal = {Journal of High Energy Physics},
  volume = {2004},
  number = {11},
  eprint = {hep-th/0408067},
  pages = {018--018},
  issn = {1029-8479},
  doi = {10.1088/1126-6708/2004/11/018},
  urldate = {2016-11-24},
  archiveprefix = {arxiv}
}

@article{Moeller:2007:ClosedBosonicString-1,
  title = {Closed {{Bosonic String Field Theory}} at {{Quintic Order}}: {{Five-Tachyon Contact Term}} and {{Dilaton Theorem}}},
  shorttitle = {Closed {{Bosonic String Field Theory}} at {{Quintic Order}}},
  author = {Moeller, Nicolas},
  year = {2007},
  month = mar,
  journal = {Journal of High Energy Physics},
  volume = {2007},
  number = {03},
  eprint = {hep-th/0609209},
  pages = {043--043},
  issn = {1029-8479},
  doi = {10.1088/1126-6708/2007/03/043},
  urldate = {2016-11-24},
  archiveprefix = {arxiv}
}

@article{Moeller:2007:ClosedBosonicString-2,
  title = {Closed {{Bosonic String Field Theory}} at {{Quintic Order II}}: {{Marginal Deformations}} and {{Effective Potential}}},
  shorttitle = {Closed {{Bosonic String Field Theory}} at {{Quintic Order II}}},
  author = {Moeller, Nicolas},
  year = {2007},
  month = sep,
  journal = {Journal of High Energy Physics},
  volume = {2007},
  number = {09},
  eprint = {0705.2102},
  pages = {118--118},
  issn = {1029-8479},
  doi = {10.1088/1126-6708/2007/09/118},
  urldate = {2016-11-24},
  archiveprefix = {arxiv}
}

@article{Moeller:2007:NonperturbativeClosedString,
  title = {The Nonperturbative Closed String Tachyon Vacuum to High Level},
  author = {Moeller, Nicolas and Yang, Haitang},
  year = {2007},
  month = apr,
  journal = {Journal of High Energy Physics},
  volume = {2007},
  number = {04},
  eprint = {hep-th/0609208},
  pages = {009--009},
  issn = {1029-8479},
  doi = {10.1088/1126-6708/2007/04/009},
  urldate = {2016-11-24},
  archiveprefix = {arxiv}
}

@article{Moeller:2008:TachyonLumpClosed,
  title = {A Tachyon Lump in Closed String Field Theory},
  author = {Moeller, Nicolas},
  year = {2008},
  month = sep,
  journal = {Journal of High Energy Physics},
  volume = {2008},
  number = {09},
  eprint = {0804.0697},
  pages = {056--056},
  issn = {1029-8479},
  doi = {10.1088/1126-6708/2008/09/056},
  urldate = {2017-12-05},
  archiveprefix = {arxiv}
}

@article{Moosavian:2019:HyperbolicGeometryClosed-1,
  title = {Hyperbolic {{Geometry}} and {{Closed Bosonic String Field Theory I}}: {{The String Vertices Via Hyperbolic Riemann Surfaces}}},
  author = {Moosavian, Seyed Faroogh and Pius, Roji},
  year = {2019},
  month = aug,
  journal = {Journal of High Energy Physics},
  volume = {1908},
  number = {arXiv:1706.07366},
  eprint = {1706.07366},
  pages = {157},
  doi = {10.1007/JHEP08(2019)157},
  urldate = {2020-04-03},
  archiveprefix = {arxiv},
  langid = {english}
}

@article{Moosavian:2019:HyperbolicGeometryClosed-2,
  title = {Hyperbolic {{Geometry}} and {{Closed Bosonic String Field Theory II}}: {{The Rules}} for {{Evaluating}} the {{Quantum BV Master Action}}},
  author = {Moosavian, Seyed Faroogh and Pius, Roji},
  year = {2019},
  month = aug,
  journal = {Journal of High Energy Physics},
  volume = {1908},
  number = {arXiv:1708.04977},
  eprint = {1708.04977},
  pages = {177},
  doi = {10.1007/JHEP08(2019)177},
  urldate = {2020-04-03},
  archiveprefix = {arxiv},
  langid = {english}
}

@article{Moosavian:2020:HyperbolicGeometrySuperstring,
  title = {Hyperbolic {{Geometry}} of {{Superstring Perturbation Theory}}},
  author = {Moosavian, Seyed Faroogh and Pius, Roji},
  year = {2020},
  month = jun,
  journal = {Fortschritte der Physik},
  volume = {68},
  number = {6},
  eprint = {1703.10563},
  primaryclass = {gr-qc, physics:hep-th, physics:math-ph},
  pages = {1900078},
  issn = {0015-8208, 1521-3978},
  doi = {10.1002/prop.201900078},
  urldate = {2022-12-13},
  archiveprefix = {arxiv}
}

@article{Pius:2014:MassRenormalizationStringGeneral,
  title = {Mass {{Renormalization}} in {{String Theory}}: {{General States}}},
  shorttitle = {Mass {{Renormalization}} in {{String Theory}}},
  author = {Pius, Roji and Rudra, Arnab and Sen, Ashoke},
  year = {2014},
  month = jul,
  journal = {Journal of High Energy Physics},
  volume = {2014},
  number = {7},
  eprint = {1401.7014},
  issn = {1029-8479},
  doi = {10.1007/JHEP07(2014)062},
  urldate = {2015-07-14},
  archiveprefix = {arxiv}
}

@article{Pius:2014:MassRenormalizationStringSpecial,
  title = {Mass {{Renormalization}} in {{String Theory}}: {{Special States}}},
  shorttitle = {Mass {{Renormalization}} in {{String Theory}}},
  author = {Pius, Roji and Rudra, Arnab and Sen, Ashoke},
  year = {2014},
  month = jul,
  journal = {Journal of High Energy Physics},
  volume = {2014},
  number = {7},
  eprint = {1311.1257},
  issn = {1029-8479},
  doi = {10.1007/JHEP07(2014)058},
  urldate = {2015-07-14},
  archiveprefix = {arxiv}
}

@book{Polchinski:2005:StringTheory-1,
  title = {String {{Theory}}: {{Volume}} 1, {{An Introduction}} to the {{Bosonic String}}},
  shorttitle = {String {{Theory}}},
  author = {Polchinski, Joseph},
  year = {2005},
  month = jun,
  publisher = {{Cambridge University Press}},
  isbn = {0-521-67227-9}
}

@misc{Scheinpflug:2023:ClosedStringTachyon,
  title = {Closed String Tachyon Condensation Revisited},
  author = {Scheinpflug, Jaroslav and Schnabl, Martin},
  year = {2023},
  month = aug,
  number = {arXiv:2308.16142},
  eprint = {2308.16142},
  primaryclass = {hep-th},
  publisher = {{arXiv}},
  doi = {10.48550/arXiv.2308.16142},
  urldate = {2023-08-31},
  archiveprefix = {arxiv}
}

@misc{Schnabl:2023:OpenStringField,
  title = {Open String Field Theory with Stubs},
  author = {Schnabl, Martin and Stettinger, Georg},
  year = {2023},
  month = mar,
  number = {arXiv:2301.13182},
  eprint = {2301.13182},
  primaryclass = {hep-th},
  publisher = {{arXiv}},
  doi = {10.48550/arXiv.2301.13182},
  urldate = {2023-04-05},
  archiveprefix = {arxiv}
}

@article{Sen:1994:ProofLocalBackground,
  title = {A {{Proof}} of {{Local Background Independence}} of {{Classical Closed String Field Theory}}},
  author = {Sen, Ashoke and Zwiebach, Barton},
  year = {1994},
  month = feb,
  journal = {Nuclear Physics B},
  volume = {414},
  number = {3},
  eprint = {hep-th/9307088},
  pages = {649--711},
  issn = {05503213},
  doi = {10.1016/0550-3213(94)90258-5},
  urldate = {2016-11-14},
  archiveprefix = {arxiv}
}

@article{Sen:1994:QuantumBackgroundIndependence,
  title = {Quantum {{Background Independence}} of {{Closed String Field Theory}}},
  author = {Sen, Ashoke and Zwiebach, Barton},
  year = {1994},
  month = jul,
  journal = {Nuclear Physics B},
  volume = {423},
  number = {2-3},
  eprint = {hep-th/9311009},
  pages = {580--630},
  issn = {05503213},
  doi = {10.1016/0550-3213(94)90145-7},
  urldate = {2016-11-14},
  archiveprefix = {arxiv}
}

@article{Sen:1996:BackgroundIndependentAlgebraic,
  title = {Background {{Independent Algebraic Structures}} in {{Closed String Field Theory}}},
  author = {Sen, Ashoke and Zwiebach, Barton},
  year = {1996},
  month = apr,
  journal = {Communications in Mathematical Physics},
  volume = {177},
  number = {2},
  eprint = {hep-th/9408053},
  pages = {305--326},
  issn = {0010-3616, 1432-0916},
  doi = {10.1007/BF02101895},
  urldate = {2016-11-14},
  archiveprefix = {arxiv}
}

@article{Sen:2015:OffshellAmplitudesSuperstring,
  title = {Off-Shell {{Amplitudes}} in {{Superstring Theory}}},
  author = {Sen, Ashoke},
  year = {2015},
  month = apr,
  journal = {Fortschritte der Physik},
  volume = {63},
  number = {3-4},
  eprint = {1408.0571},
  pages = {149--188},
  issn = {00158208},
  doi = {10.1002/prop.201500002},
  urldate = {2016-04-17},
  archiveprefix = {arxiv}
}

@article{Sen:2018:BackgroundIndependenceClosed,
  ids = {Sen:2017:BackgroundIndependenceClosed},
  title = {Background {{Independence}} of {{Closed Superstring Field Theory}}},
  author = {Sen, Ashoke},
  year = {2018},
  month = feb,
  journal = {Journal of High Energy Physics},
  volume = {2018},
  number = {2},
  eprint = {1711.08468},
  pages = {155},
  issn = {1029-8479},
  doi = {10.1007/JHEP02(2018)155},
  urldate = {2020-04-03},
  archiveprefix = {arxiv}
}

@article{Sen:2020:DinstantonPerturbationTheory,
  title = {D-Instanton {{Perturbation Theory}}},
  author = {Sen, Ashoke},
  year = {2020},
  month = may,
  eprint = {2002.04043},
  url = {http://arxiv.org/abs/2002.04043},
  urldate = {2020-05-30},
  archiveprefix = {arxiv}
}

@article{Sen:2021:DinstantonsStringField,
  title = {D-Instantons, {{String Field Theory}} and {{Two Dimensional String Theory}}},
  author = {Sen, Ashoke},
  year = {2021},
  month = nov,
  journal = {Journal of High Energy Physics},
  volume = {2021},
  number = {11},
  eprint = {2012.11624},
  primaryclass = {hep-th},
  pages = {61},
  issn = {1029-8479},
  doi = {10.1007/JHEP11(2021)061},
  urldate = {2023-02-27},
  archiveprefix = {arxiv}
}

@article{Sonoda:1989:HermiticityCPTString,
  title = {Hermiticity and {{CPT}} in String Theory},
  author = {Sonoda, Hidenori},
  year = {1989},
  month = oct,
  journal = {Nuclear Physics B},
  volume = {326},
  number = {1},
  pages = {135--161},
  issn = {0550-3213},
  doi = {10.1016/0550-3213(89)90437-9},
  urldate = {2019-04-16}
}

@article{Sonoda:1990:CovariantClosedString,
  title = {Covariant Closed String Theory Cannot Be Cubic},
  author = {Sonoda, Hidenori and Zwiebach, Barton},
  year = {1990},
  month = may,
  journal = {Nuclear Physics B},
  volume = {336},
  number = {2},
  pages = {185--221},
  issn = {0550-3213},
  doi = {10.1016/0550-3213(90)90108-P},
  urldate = {2018-03-25}
}

@article{Stratonovich:1957:MethodCalculatingQuantum,
  title = {On a {{Method}} of {{Calculating Quantum Distribution Functions}}},
  author = {Stratonovich, R. L.},
  year = {1957},
  month = jul,
  journal = {Soviet Physics Doklady},
  volume = {2},
  pages = {416},
  urldate = {2018-08-10}
}

@article{Witten:1986:NoncommutativeGeometryString,
  title = {Non-Commutative Geometry and String Field Theory},
  author = {Witten, Edward},
  year = {1986},
  month = may,
  journal = {Nuclear Physics B},
  volume = {268},
  number = {2},
  pages = {253--294},
  issn = {0550-3213},
  doi = {10.1016/0550-3213(86)90155-0},
  urldate = {2013-12-19}
}

@article{Yang:2005:ClosedStringTachyon,
  title = {A {{Closed String Tachyon Vacuum}}?},
  author = {Yang, Haitang and Zwiebach, Barton},
  year = {2005},
  month = sep,
  journal = {Journal of High Energy Physics},
  volume = {2005},
  number = {09},
  eprint = {hep-th/0506077},
  pages = {054--054},
  issn = {1029-8479},
  doi = {10.1088/1126-6708/2005/09/054},
  urldate = {2016-11-14},
  archiveprefix = {arxiv}
}

@article{Yang:2005:DilatonDeformationsClosed,
  title = {Dilaton {{Deformations}} in {{Closed String Field Theory}}},
  author = {Yang, Haitang and Zwiebach, Barton},
  year = {2005},
  month = may,
  journal = {Journal of High Energy Physics},
  volume = {2005},
  number = {05},
  eprint = {hep-th/0502161},
  pages = {032--032},
  issn = {1029-8479},
  doi = {10.1088/1126-6708/2005/05/032},
  urldate = {2016-11-24},
  archiveprefix = {arxiv}
}

@article{Yang:2005:TestingClosedString,
  title = {Testing {{Closed String Field Theory}} with {{Marginal Fields}}},
  author = {Yang, Haitang and Zwiebach, Barton},
  year = {2005},
  month = jun,
  journal = {Journal of High Energy Physics},
  volume = {2005},
  number = {06},
  eprint = {hep-th/0501142},
  pages = {038--038},
  issn = {1029-8479},
  doi = {10.1088/1126-6708/2005/06/038},
  urldate = {2016-11-24},
  archiveprefix = {arxiv}
}

@article{Zwiebach:1988:ConstraintsCovariantTheories,
  title = {Constraints on Covariant Theories for Closed String Fields},
  author = {Zwiebach, Barton},
  year = {1988},
  month = aug,
  journal = {Annals of Physics},
  volume = {186},
  number = {1},
  pages = {111--140},
  issn = {0003-4916},
  doi = {10.1016/S0003-4916(88)80019-8},
  urldate = {2018-07-06}
}

@article{Zwiebach:1993:ClosedStringField,
  title = {Closed {{String Field Theory}}: {{Quantum Action}} and the {{BV Master Equation}}},
  shorttitle = {Closed {{String Field Theory}}},
  author = {Zwiebach, Barton},
  year = {1993},
  month = jan,
  journal = {Nuclear Physics B},
  volume = {390},
  number = {1},
  eprint = {hep-th/9206084},
  pages = {33--152},
  issn = {05503213},
  doi = {10.1016/0550-3213(93)90388-6},
  urldate = {2016-01-21},
  archiveprefix = {arxiv}
}

\end{document}